\documentclass[twocolumn]{aa}  

\usepackage{graphicx}
\usepackage{txfonts}
\usepackage{xspace}
\usepackage{mathrsfs}
\newcommand{\nnhp}{$\rm N_2H^+$\xspace}
\newcommand{\hhdp}{$\rm H_2D^+$\xspace}
\newcommand{\nndp}{$\rm N_2D^+$\xspace}
\newcommand{\hcop}{$\rm HCO^+$\xspace}
\newcommand{\dcop}{$\rm DCO^+$\xspace}
\newcommand{\hcdop}{$\rm HC^{18}O^+$\xspace}
\newcommand{\kms}{$\rm km \, s^{-1}$\xspace}

\newcommand{\xab}{$f_\mathrm{corr}$\xspace}
\newcommand{\modh}{$\mathscr{H}$\xspace}
\newcommand{\modf}{$\mathscr{F}$\xspace}
\newcommand{\modl}{$\mathscr{L}$\xspace}
\newcommand{\crir}{$\zeta_2$\xspace}

\begin{document}

   \title{The cosmic-ray ionisation rate in the pre-stellar core L1544\thanks{This work is based on observations carried out with the IRAM 30m telescope. IRAM is supported by INSU/CNRS (France), MPG (Germany) and IGN (Spain).}}

\titlerunning{CR ionisation rate in L1544}

   \author{E. Redaelli
          \inst{1}
          \and
        O. Sipil\"a\inst{1}
        \and
        M. Padovani \inst{2}
        \and
        P. Caselli \inst{1}
        \and
        D. Galli \inst{2}
        \and
        A. V. Ivlev \inst{1}
          }

   \institute{Centre for Astrochemical Studies, Max-Planck-Institut f\"ur extraterrestrische Physik, Gie{\ss}enbachstra{\ss}e 1, 85749 Garching bei M\"unchen,
Germany\\
              \email{eredaelli@mpe.mpg.de}
         \and
            INAF--Osservatorio Astrofisico di Arcetri, Largo E. Fermi 5, 50125 Firenze, Italy             }

   \date{...;...}

  \abstract
   {Cosmic rays (CRs), energetic particles composed mainly by protons and electrons, play an important role in the chemistry and dynamics of the interstellar medium. In dense environments, they represent the main ionising agent, hence driving the rich chemistry of molecular ions. Furthermore, they determine the ionisation fraction, which regulates the degree of coupling between the gas and the interstellar magnetic fields, and the heating of the gas. Estimates of the CR ionisation rate of molecular hydrogen (\crir) span several orders of magnitude, depending on the targeted sources and on the used method.}
   {Recent theoretical models have characterised the CR attenuation with increasing density. We aim to test these models for the attenuation of CRs in the low-mass pre-stellar core L1544. }
   {We use a state-of-the-art gas-grain chemical model, which accepts the CR ionisation rate profile as input, to predict the abundance profiles of four ions: \nnhp, \nndp, \hcdop, and \dcop. Non-local thermodynamic equilibrium radiative transfer simulations are run to produce synthetic spectra based on the derived abundances. These are compared with observations obtained with the Institut de Radioastronomie Millim\'etrique (IRAM) 30m telescope.}
   {Our results indicate that a model with high \crir ($> 10^{-16} \rm \, s^{-1}$) is excluded by the observations. Also the model with the standard $\zeta_2 = 1.3 \times 10^{-17} \rm \, s^{-1}$ produces a worse agreement with respect to the attenuation model based on Voyager observations, which is characterised by an average $\langle \text{\crir} \rangle = 3 \times 10^{-17} \rm \, s^{-1}$ at the column densities typical of L1544. The single-dish data, however, are not sensitive to the attenuation of the CR profile, which changes only by a factor of two in the range of column densities spanned by the core model ($N = 2-50\times \rm 10^{21} \, cm^{-2}$). Interferometric observations at higher spatial resolution, combined with observations of transitions with lower critical density ---hence tracing the low-density envelope--- are needed to observe a decrease of the CR ionisation rate with density. 
}
   {}


   \maketitle
%

\section{Introduction\label{Intro}}
Cosmic rays (CRs) are ubiquitous in the interstellar medium (ISM), and play a leading role in determining its ionisation degree. In the denser regions of molecular clouds, where the column density is $N \gtrsim 10^{21} \, \rm cm^{-2}$ (corresponding to visual extinction in magnitude $A_\mathrm{V} \gtrsim 1$), ultraviolet (UV) photons of the interstellar radiation field are efficiently absorbed and cannot penetrate. In these physical conditions, CRs become the main {ionising agent} of the gas, with important consequences. First of all, they affect the chemistry by producing ions. In particular, the first reaction is the ionisation of hydrogen molecules:
\begin{equation}
\mathrm{ CR + H_2 \rightarrow H_2^+  } + \rm e^- + {CR} \; .
\end{equation}
The ionised $\rm H_2^+$ quickly reacts with another hydrogen molecule, producing the fundamental trihydrogen cation ($\rm H_3^+$). $\rm H_3^+$ in turn is the starting point for the chain of reactions between charged and neutral species, producing several other key molecules, such as \hcop, via reaction with CO. Indirectly, CRs regulate also another important chemical process in the cold phases of the ISM: deuteration. Arguably the most important deuterated species in the gas phase is \hhdp, which is produced by isotopic exchange reactions between $\rm H_3^+ $ and $\rm HD$ \citep{Dalgarno84}. Moreover, CRs produce $\rm He^+$, which can contribute to the destruction of CO, liberating C atoms in the gas phase which can then form carbon-chain molecules (see e.g. \citealt{Ruffle99}), and to the formation of N$^+$ from N$_2$, initiating the reaction chain leading to ammonia formation. \par
Furthermore, by determining the ionisation fraction of the ISM, CRs also affect its dynamical evolution. The ISM is threaded by magnetic fields (see e.g. the results of \citealt{PlanckXXXV} for the magnetic properties of molecular clouds), which exert their influence directly on charged particles. These transfer momentum to the neutral component through drag forces (collisions). The ionisation fraction therefore determines the degree of coupling between the gas and the magnetic field. In particular, the fraction of free electrons determines the timescale of ambipolar diffusion, the process of drift of neutral matter across the field lines which is the proposed mechanism that allows the gravitational collapse of magnetised cores \citep{Mouschovias76}. \par
Despite its importance, observational estimates of the CR ionisation rate are still scarce, uncertain, and scattered over a wide range of values. The most direct and reliable way to determine it is represented by direct observations of $\rm H_3^+$. These, however, are possible only in  diffuse gas via infrared spectroscopy. Lacking a permanent electric dipole, in fact, $\rm H_3^+$ is not observable in emission {in the interstellar gas} at high densities. In {dense} clouds, \cite{VanDerTak00} used $\rm H_3^+$ absorption lines in combination with $\rm H^{13}CO^+$ rotational lines towards background massive protostars, and they determined a CR ionisation rate of $\approx 3\times 10^{-17} \, \rm s^{-1}$. More recently, \cite{Indriolo12} ---expanding the sample from \cite{Indriolo07}--- reported 21 detections toward diffuse clouds, determining CR ionisation rate per hydrogen molecule (\crir) in the range $[1.7 -10.6]\times 10^{-16} \rm \, s^{-1}$.  \cite{Neufeld17} derived a CR ionisation rate per hydrogen atom ($\zeta_1$) of the order of a few $10^{-16} \rm \, s^{-1}$ in the Galactic disc, using $\rm H_3^+$ and other ionised tracers. Their data support a steep dependency of \crir on the gas column density ($\zeta_2 \propto N^{-1}$).\par
In {regions so dense that no background source is visible at the near infrared wavelengths of the $\rm H_3^+$ lines, no $\rm H_3^+$ absorption can be observed. In these conditions,} a commonly used approach is to compare the abundances of molecular tracers sensitive to the ionisation fraction (such as \hcop) with the results of chemical models \citep[see e.g.][]{Black78}. This method was later used for instance by \cite{Caselli98} towards several pre-stellar and proto-stellar cores, using observations of \dcop, \hcop, and CO. The results span the range $10^{-18} - 10^{-16} \, \rm s^{-1}$. \cite{Bovino20} proposed a new analytical method to assess \crir using observations of CO, of the deuterium fraction (for instance from \hcop and \dcop), and of ortho-$\rm H_2D^+$. The method was applied by \cite{Sabatini20} to a sample of massive clumps, finding $\text{\crir} \approx [0.7- 6] \times 10^{-17}\rm \, s^{-1} $. \par
From the theoretical point of view, {one of the first studies} on CR ionisation was performed by \cite{Spitzer68}, who computed the CR ionisation rate of atomic hydrogen, yielding $\zeta_1 \approx 6.8 \times 10^{-18} \rm \, s^{-1} $. This can be transformed into \crir using the relation $\zeta_2 = (2.3/1.5) \times \zeta_1$ \citep{Glassgold74}, obtaining the value $\text{\crir} \approx 10^{-17}\rm \, s^{-1} $ which is still considered the standard value for molecular clouds. However, the assumption that \crir is independent on the ISM density is incorrect, as also shown by the different observational results. CRs lose energy in their interactions with the ambient medium, which modifies their spectrum. These losses hence cause an attenuation of \crir, which decreases with gas column density ($N$). This process has been studied in detail by \cite{Padovani09}, who computed the relation between the CR ionisation rate and the gas column density using: 
\begin{equation}
\label{zeta}
\zeta_2(N) = 4\pi[1+\Phi(N) ] \sum_k  \int_{ E_{\rm ion}}^{\infty} j_k(E,N) \sigma_k^\mathrm{ion}(E) \, dE \; ,
\end{equation}
where $k$ indicates the primary CR species (e.g. protons, electrons, positrons, and heavier nuclei),  $E_{\rm ion} = 15.44\, \mathrm{eV}$ is the energy threshold for H$_2$ ionisation, $\sigma_k^\mathrm{ion}(E)$ is the ionisation cross section of H$_2$ by the species $k$ \citep{Rudd92, Kim00}, and $ j_k(E,N)$ is its local energy spectrum
, which depends on $N$ due to energy losses. Secondary CR electrons, produced due to the gas ionisation by the primary CR species, lead to additional H$_2$ ionisation. Their contribution is described by $\Phi(N)$, which is the ratio between the secondary to primary H$_2$ ionisation rates (see \citealt{Ivlev21}). {The local CR spectrum is commonly calculated using the continuous slowing-down approximation (see, e.g., \citealt{Padovani09}), which assumes that CRs propagate freely along the local magnetic field lines. As was shown by \cite{Silsbee18}, the results are generally applicable to arbitrary magnetic field configurations, including converging field lines: as long as the field strength increases monotonically along the field line, the effects of magnetic mirroring and focusing practically cancel each other. The column density $N$ in this case is measured along the field lines.}

\par

\cite{Padovani18} (hereafter PI18) further expanded {the work of \cite{Padovani09}} considering the regime of higher column densities, and using the most recent data from the Voyager missions to constrain the CR interstellar flux \citep{Cummings16, Stone19}. Similarly to \cite{Ivlev15}, PI18 considered two models for the interstellar flux of protons: the first, called the ``high'' model, is characterised by $\zeta_2 \gtrsim 10^{-16} \rm \, s^{-1}  $ up to $N \approx \text{ a few } \times 10^{23} \rm \, cm^{-2}$, and reproduces the high \crir measured in the diffuse medium. The ``low'' model, which instead implies $\zeta_2 \approx \text{ a few } 10^{-17} \rm \, s^{-1}$, is the one that fits the Voyager 1 and 2 data. It is important to notice that the spectrum derived by the Voyager missions could represent a lower limit, due to \textit{i)} persisting contamination from CRs from inside the heliosphere \citep{Scherer08}, and to \textit{ii)} the possible influence of local sources, which make the observed spectrum deviate from the average Galactic one \citep{Phan21}. 
\par
In this work, we investigate the CR ionisation rate in the prototypical pre-stellar core L1544, combining CR propagation models from PI18 with a state-of-the-art chemical {model}, and comparing the simulated spectra of several rotational lines of molecular ions with recent observations. L1544 is situated in the Taurus molecular cloud at a distance of $ \approx 135 \,$pc \citep{Schlafly14}, and it shows bright
continuum emission at mm and sub-mm wavelengths \citep{WardThompson99, Doty04, Spezzano16}. Its physical structure has been modelled by means of a quasi-equilibrium Bonnor-Ebert sphere using \nnhp and carbon monoxide spectra \citep[e.g.][]{Keto15}. This model predicts low temperatures ($T\approx 6\, \rm K$) and high volume densities ($n \gtrsim 10^6\, \rm cm^{-3}$) within the central 2000$\,$AU, in agreement with  observations \citep{Crapsi07, ChaconTanarro17, Caselli19}.
\par
 In a recent work \citep[][hearafter RB19]{Redaelli19}, we combined this physical model with a gas-grain chemical model to investigate the deuteration maps of {\hcop and \nnhp (diazenylium)} across the source. We now perform a similar analysis, focusing this time on \crir. In Sect. \ref{ModObs} we present the chemical and physical models used, and the resulting abundance profiles are discussed in Sect. \ref{Results}. The analysis of the radiative transfer results for the different cases are reported in Sect. \ref{Analysis}. Section \ref{Concl} summarises the main findings and conclusions of this work. 

\section{Model description and Observations\label{ModObs}}
\subsection{Physical model}
The physical model of L1544 consists of a one-dimensional, quasi-equilibrium Bonnor-Ebert sphere, developed by \cite{Keto15} fitting $\rm C^{18}O$, $\rm H_2O$, and \nnhp lines. It comprises the gas temperature, dust temperature, volume density, and infall velocity profile, and it has been successfully used to model several molecules at the dust peak of L1544 \citep{Bizzocchi13, Redaelli18, Redaelli19}. In the latest paper, we modelled the {\nnhp and \hcop} isotopologues using a modified infall velocity profile (scaled up by a factor $1.75$), which was originally introduced by \cite{Bizzocchi13} to reproduce the \nnhp (1-0) transition with a constant abundance profile. We realised that this enhancement of the infall velocity leads to incorrect spectral profiles (e.g. strong double-peak profiles in \nndp), which are inconsistent with the observations. For this work we therefore use the original velocity profile of \cite{Keto15}.  The volume density profile  and the corresponding integrated column density profile are shown in the left {and central panels} of Fig. \ref{CRIR}.

\subsection{Chemical models}
\begin{figure*}[!h]
\centering
\includegraphics[width=.9\textwidth]{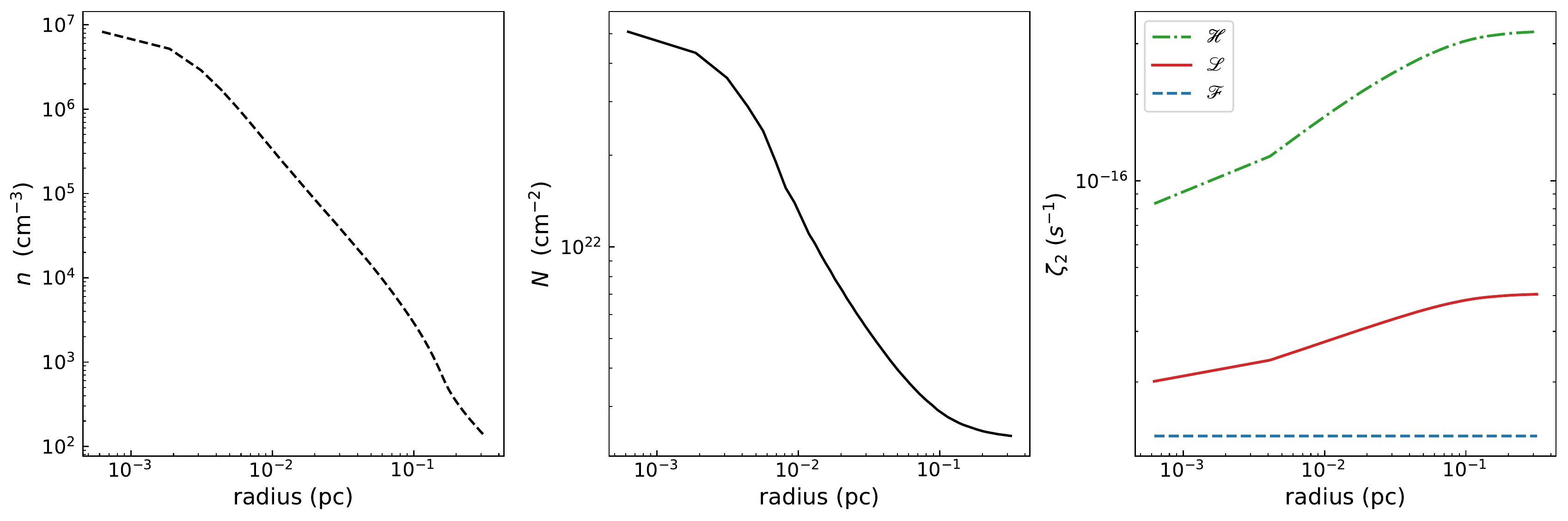}
\caption{{\textit{Left panel:} H$_2$ volume density profile ($n$) in the physical model of L1544. \textit{Central panel:}  column density profile.} \textit{Right panel:} The \crir profiles for the \modf (dashed blue curve), \modl (solid red), and \modh (dashed-dotted green) models. The profiles for the models \modl and \modh  PI18 have been derived integrating the {volume} density profile of the core model to obtain the column density profile, and then using the relation between $\text{\crir}$ and $N$. \label{CRIR}}
\end{figure*}

The physical model of L1544 is coupled to the gas-grain chemical model presented in \cite{Sipila15,Sipila15b,Sipila19b}. The core model is divided into concentric shells, and the evolution of the chemistry is then solved in each shell separately. The combination of the result from all the shells at a common time step gives the radial profile of the molecular abundance. {The radial grid consists of 35 points covering the whole core size, assumed to be $0.32 \, \rm pc$ (see Fig. \ref{CRIR}), with a resolution that ranges from $3.5 \times 10^{-3} \, \rm pc$ in the central parts to $2\times 10^{-2} \, \rm pc$ in the outskirts of the core model.} The initial conditions are summarised in Table \ref{InitialCond}. The dust population consists of grains of $0.1\, \rm \mu m$ of radius and a grain material density of $\rho_\mathrm{g} = 2.5 \, \rm g \, cm^{-3}$. The effect of the embedding cloud is simulated by assuming an external visual extinction of $A_\mathrm{V} = 2 \, \rm mag$\footnote{RB19 tested $A_\mathrm{V}$ values of  $1, 2, \text{ and } 5 \, \rm mag$, and showed that this parameter does not affect significantly the synthetic spectra of the targeted species, in particular of \nnhp isotopologues.}, which attenuates the external UV field. We test four different time steps: $t = 10^5$, $5\times 10^5$, $10^6$, and  $5\times 10^6\, \rm yr$.\par

\begin{table}[!h]
\renewcommand{\arraystretch}{1.4}
\caption{Initial abundances with respect to {the total number density of hydrogen nuclei} used in the chemical model.}
\centering
  \begin{tabular}{cc}
  \hline
  Species   & Initial abundance\\ 
  \hline \hline
  H$_{2}$   & 0.5\\
  He    & 9.00$\times$10$^{-2}$\\
  C$^{+}$& 1.20$\times$10$^{-4}$\\
  N    & 7.60$\times$10$^{-5}$ \\
  O & 2.56$\times$10$^{-4}$\\
   S$^{+}$   & 8.00$\times$10$^{-8}$\\
   Si$^{+}$        &8.00$\times$10$^{-9}$ \\
   Na$^{+}$         &2.00$\times$10$^{-9}$ \\
  Mg$^{+}$         & 7.00$\times$10$^{-9}$\\
Fe$^{+}$     & 3.00$\times$10$^{-9}$\\
P$^{+}$      &2.00$\times$10$^{-10}$ \\
   Cl$^{+}$      & 1.00$\times$10$^{-9}$\\
 F    & 2.00$\times$10$^{-9}$\\
H$_2$ ortho-to-para ratio & 10$^{-3}$\\
      \hline
  \label{InitialCond}
  \end{tabular}
\end{table}

The chemical code takes as input the CR ionisation-rate $\text{\crir}$, which can be either constant, or can have a radial dependence. For the purposes of this paper, we have tested three models:
\begin{enumerate}
\item The fiducial model (\modf), with a constant value $\text{\crir} = 1.3 \times 10^{-17} \, \rm s^{-1}$;
\item The high model (\modh) from PI18;
\item The low model (\modl) from PI18.
\end{enumerate}
The first model is similar to that used in RB19. 
In order to implement models \modl and \modh, we have used the parametrisation presented in Appendix F of PI18 to compute the \crir profile from the column density profile. The radial profiles of the three \crir models are presented in the right panel of Fig. \ref{CRIR}. {In models \modl and \modh, the \crir profile is computed assuming a column density value corresponding to $A_\mathrm{V} = 2$ at the edge of the core.} The three models are well separated in the range of \crir values that they span, and can be ordered by increasing \crir as \modf, \modl, and then \modh.
\par

{In our approach, we assume for simplicity that CRs propagate radially along the field lines in a spherically symmetric geometry, but we make no assumption on the field geometry (except the one that the field strength does not have multiple minima along the field lines, see Sect. \ref{Intro}). Accordingly, we compute the column density for CR attenuation by integrating radially the density from the cloud's boundary to the point of interest. This is the minimum possible  column density that characterises the CRs attenuation at any interior point of the cloud. In general, the attenuation of CRs propagating along magnetic field lines will be larger than this minimum value, by a factor of the order of a few times that depends on the actual field geometry (see \citealt{Padovani13} for an accurate calculation of the column density in a realistic cloud's magnetic field and density profile). We note that at small scales ($< 10^{-3} \rm \, pc = 200 \, AU$), where the field lines can become entangled and present complex morphology with local minima of the field strength (see again \citealt{Padovani13}), the magnetic mirroring and focusing may no longer balance each other, and then calculating the local ionisation rate becomes a nontrivial problem \citep{Silsbee18}. However, here we investigate processes occurring at substantially larger physical scales ($> 10^{-2}\, \rm pc = 2000 \, AU$), and therefore we can safely employ the results of PI18.}

\par
We stress here that the models \modf, \modl, and \modh differ in the shape of the \crir profile, whilst the rest of the physical structure (in terms of velocity, temperature, and density) remains the same. This approach is not self-consistent, since the ionisation rate affects the temperature of the gas. A complete physical model should be computed self-consistently, which is however beyond the scope of the present paper. We do not expect this effect to be significant in the dense gas traced by the targeted molecular transitions, as the gas at these densities ($n \gtrsim 10^5 \rm \, cm^{-3}$) is mainly cooled by dust. This cooling is efficient enough that we do not expect important changes \citep[see also][]{Ivlev19}. We however discuss this point in more detail in Appendix \ref{DifferentT}. Our chemical network does not include the chemistry of oxygen isotopes. The abundance profiles for \hcdop , which is the observed isotopologue, are derived from those of \hcop, by {dividing the latter} by the isotopic ratio $\rm ^{16}O / ^{18}O = 557$ \citep{Wilson99}.

\subsection{Observations}
The abundance profiles derived from our chemical network can be used to obtain synthetic spectra to be compared with observations. For this, we have used the data presented in RB19. Specifically, we use single-dish data obtained with the Institut de Radioastronomie Millim\'etrique (IRAM) 30m telescope of: \nnhp (1-0) and (3-2); \nndp (1-0), (2-1), and (3-2); \dcop (1-0), (2-1), and (3-2); \hcdop (1-0). The latest is preferred to the optically thick main isotopologue. All the spectra are obtained as single pointings taken at the millimetre dust peak of L1544 ($\rm R.A.(J2000)= 05^h04'17''.21$, $\rm Dec.(J2000)= 25^\circ10'42''.8$). We refer to RB19 for a complete description of the observational data. We remark however the unique features of these data, which makes them particularly suitable for the purposes of our work. First of all, with the exception of the \dcop (1-0) transition, all the data have been acquired with good spectral resolution ($\Delta V = 0.02-0.08 \, $\kms), crucial to correctly recover the complex line profiles arising from a combination of molecular depletion, self-absorption, and kinematic effects. Furthermore, the different rotational transitions of a given species have distinct critical densities, and hence trace different depths into the core. For instance, at $10\,$K the critical densities of the \nnhp (1-0) and (3-2) transitions are $n_\mathrm{c} =6 \times 10^4 \, \rm cm^{-3}$ and $n_\mathrm{c} =1.4\times 10^6 \, \rm cm^{-3}$, respectively \citep{Shirley15}. This fact, combined with the improved angular resolution at higher frequencies, allows to test reliably the radial profiles of these molecules, even though the spatial resolution of the single-dish data is limited for the low-$J$ transitions. 

\section{Results\label{Results}}
\begin{figure*}[!t]
\centering
\includegraphics[width=\textwidth]{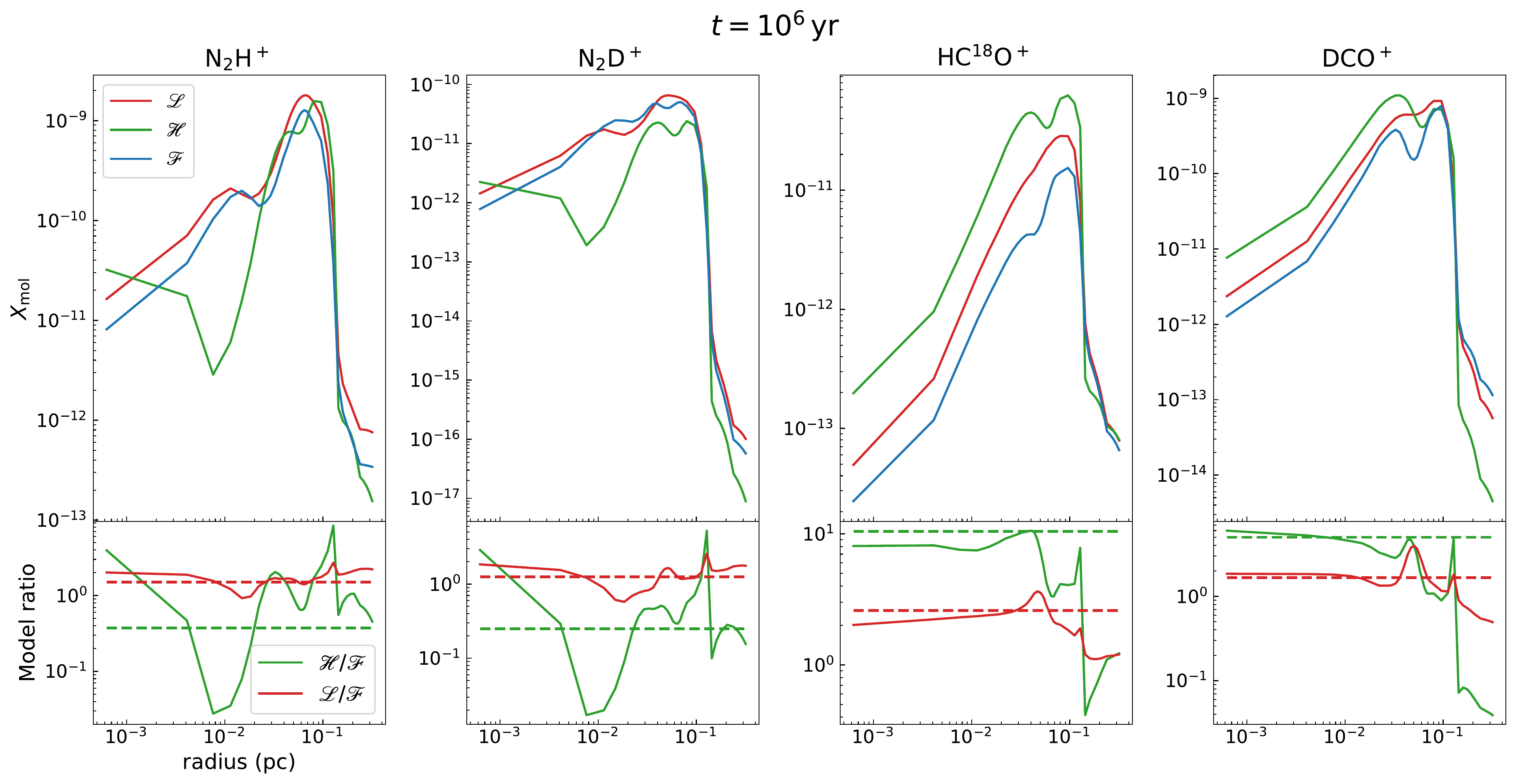}
\caption{\textit{Top panels:} Abundance profiles for the four analysed species (from left to right: \nnhp, \nndp, \hcdop, \dcop) at $ t =  10^6\, \rm yr$. The colours refer to different models used for the CR ionisation rate: model \modf (blue), model \modl (red), and model \modh (green). \textit{Bottom panels:} The green and {red} solid curves show the ratio of model \modh and model \modl to model \modf, respectively. The dashed, horizontal lines represent the inverse ratios of the corrective factors \xab used to reproduce the observed lines, as reported in Sect. \ref{mod:1e6} and Table \ref{xab_summ}. To give an example, the radiative transfer of \dcop lines requires $\text{\xab} = 0.9$ in model \modl, and $\text{\xab} = 1.5$ in model \modf; therefore the {red} solid curve \modl/\modf in the rightmost panel has to be compared with the {red} dashed line at $(0.9/1.5)^{-1}=1.67$.   \label{ChemMod}}
\end{figure*}

Figure \ref{ChemMod} shows the simulated abundance profiles for the different species of interest at $ t =  10^6\, \rm yr$. We show in Appendix \ref{AllModels} the plots for the remaining time steps. From the comparison of the models, we can derive some general conclusions. As the evolutionary time increases, \nnhp isotopologues show stronger and stronger depletion towards the core's centre. The abundance peak increases, but shifts outwards, whilst the abundance at smaller radii decreases. This effect is visible also for \dcop and \hcdop, but it is less evident because these molecules are already significantly depleted early on ($t \lesssim 10^5 \, \rm yr$) at radii smaller than $0.04\, \rm pc$. 
\par
The effect of varying \crir is species-dependent. \hcop (and as a consequence \hcdop) presents a clear, monotonic effect at every time step: the higher the value of \crir, the higher the overall abundance profile, due to the increase of the ionisation fraction. This trend is partially visible also for \dcop, but the differences in the abundance profiles of the various models are smaller. The behaviour of \nnhp isotopologues is more complex, and time-dependent. In general, we note that the higher the overall \crir, the stronger is the depletion, and the earlier this phenomenon starts showing up. 
Model \modh predicts a strong drop in the abundance of \nnhp and \nndp starting at $ t =  10^6\, \rm yr$. At $ t =  5 \times 10^6\, \rm yr$, also model \modl (which has an intermediate \crir value between models \modf and \modh) predicts an enhanced decrease of the abundances. For \nndp in particular, the high model tends to always predict the lowest abundances with respect to the other two models. 

\section{Analysis and Discussion \label{Analysis}}
\begin{figure*}[!t]
\centering
\includegraphics[width=.85\textwidth]{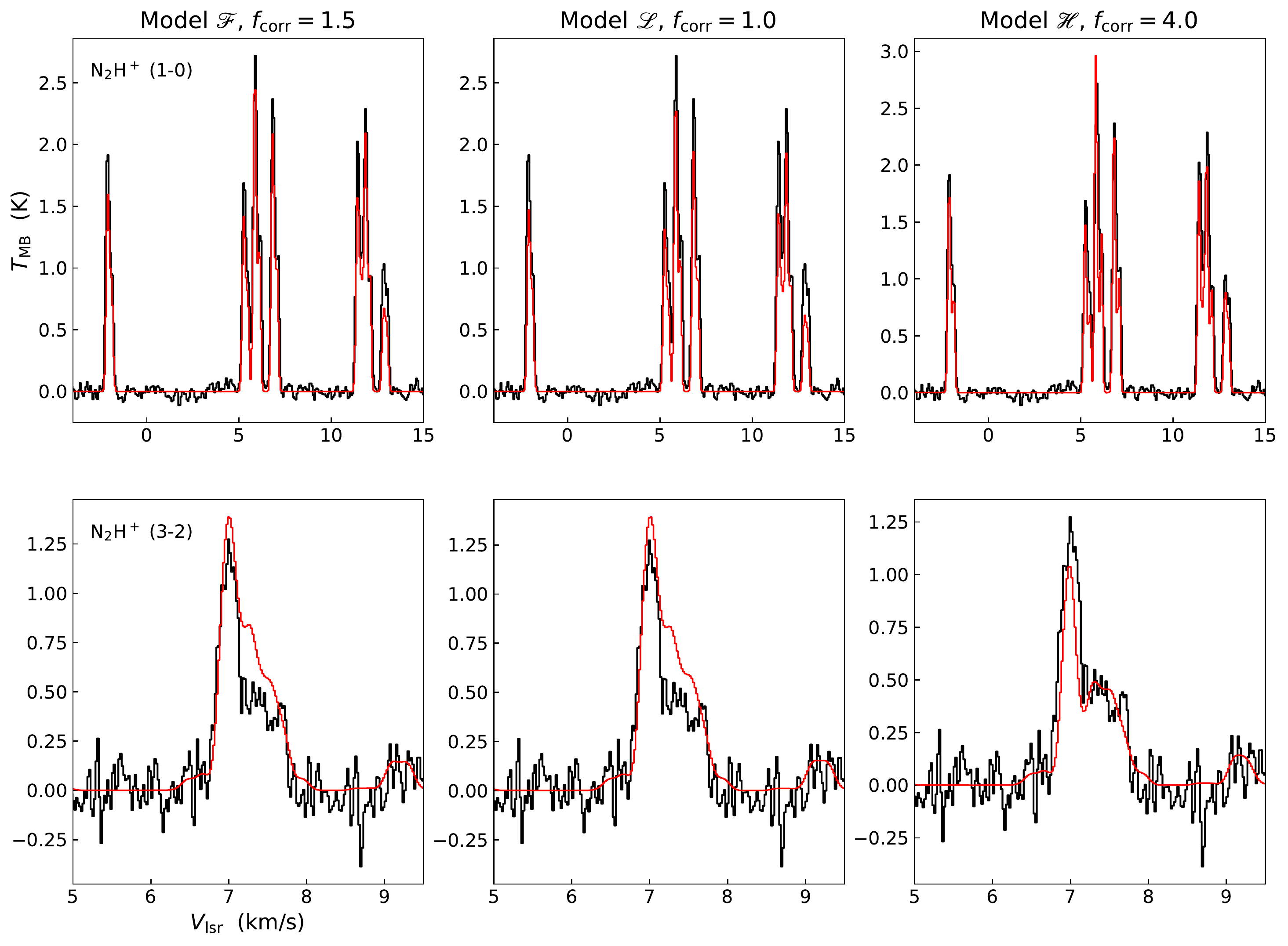}
\caption{Spectra obtained with the radiative transfer code (model, in red) overlaid to the observations (black) for \nnhp (1-0) and (3-2) lines. The columns refer to the three models: \modf in the left panels, \modl in the central panels, and \modh in the right ones. The corrective factors \xab used to modify the whole abundance profile to match the observed fluxes is indicated above the top panel of each column. All models are at $ t =  10^6\, \rm yr$. \label{Models_n2hp}}
\end{figure*}

\subsection{Non-LTE modelling at $ t =  10^6\, \rm yr$\label{mod:1e6}}
In order to produce the synthetic spectra to be compared with the observations we have used the non-local-thermodynamic-equilibrium (non-LTE) radiative transfer code MOLLIE \citep{Keto90, Keto04}. MOLLIE requires several inputs: the source's physical structure; the molecular abundance profile, obtained from the chemical models, which can be modified by a corrective factor \xab\footnote{We highlight that multiplying the abundance profile by \xab does not correspond to multiplying the fluxes modelled by radiative transfer by the same factor, due to opacity and non-LTE effects.}; the collisional coefficients of the molecular transitions (see RB19 for references); the turbulent line broadening that is summed in quadrature to the thermal one. For {this} we adopted $\sigma_\mathrm{turb}= 0.075 \, \rm $ \kms, following RB19. The code, which is particularly suited to model crowded hyperfine structures prone to selective photon trapping, requires long computational times to converge (of the order of hours using 20 cores for \nnhp and \nndp), and a full parameter-space exploration is not feasible, in particular to determine \xab. As in RB19, we therefore made several tests to determine by trial-and-error the \xab value that matches the observed line intensities for each species.  \par
{Before performing the ray-tracing to compute the synthetic spectra, MOLLIE interpolates the 1D physical model and abundance profile to a 3D array (mantaining the spherical symmetry), which consists of three nested grids that become finer going towards the core centre. Going from the largest to the smallest grid, each of them covers half of the radial range with respect to the previous one, doubling the spatial resolution. The finest grid has a cell size of $1.7 \times 10^{-3} \, \rm pc$, which corresponds to $\approx 2.6''$ at the distance of L1544. We conclude that the models produced by MOLLIE have much higher resolution than the observations. In order to allow for a proper comparison with the observed data, we then convolve the results from MOLLIE to the corresponding beam size for each transition, and we extract the spectra at the centre of the models.}
\par
The results of the chemical simulations tend to over- or under-estimate the observed fluxes, in particular when one tries to model several transitions of different species at the same time. We base the comparison among different models mainly on three parameters:
\begin{enumerate}
\item the corrective \xab required to reproduce the observed fluxes; the closer these are to unity, the better the model performs;
\label{list1}
\item the agreement with the line profiles (in term of asymmetries, double-peak features, self-absorption, and intensity ratios of the hyperfine components, if present);\label{list2}
\item the correct prediction of the intensity ratios of different rotational transitions for the same species. This is particularly important, because due to the different excitation conditions of the distinct rotational lines, their intensity ratios probe the level of depletion of the species at high densities.
\label{list3}
\end{enumerate}

\par

\begin{figure*}[!h]
\centering
\includegraphics[width=.85\textwidth]{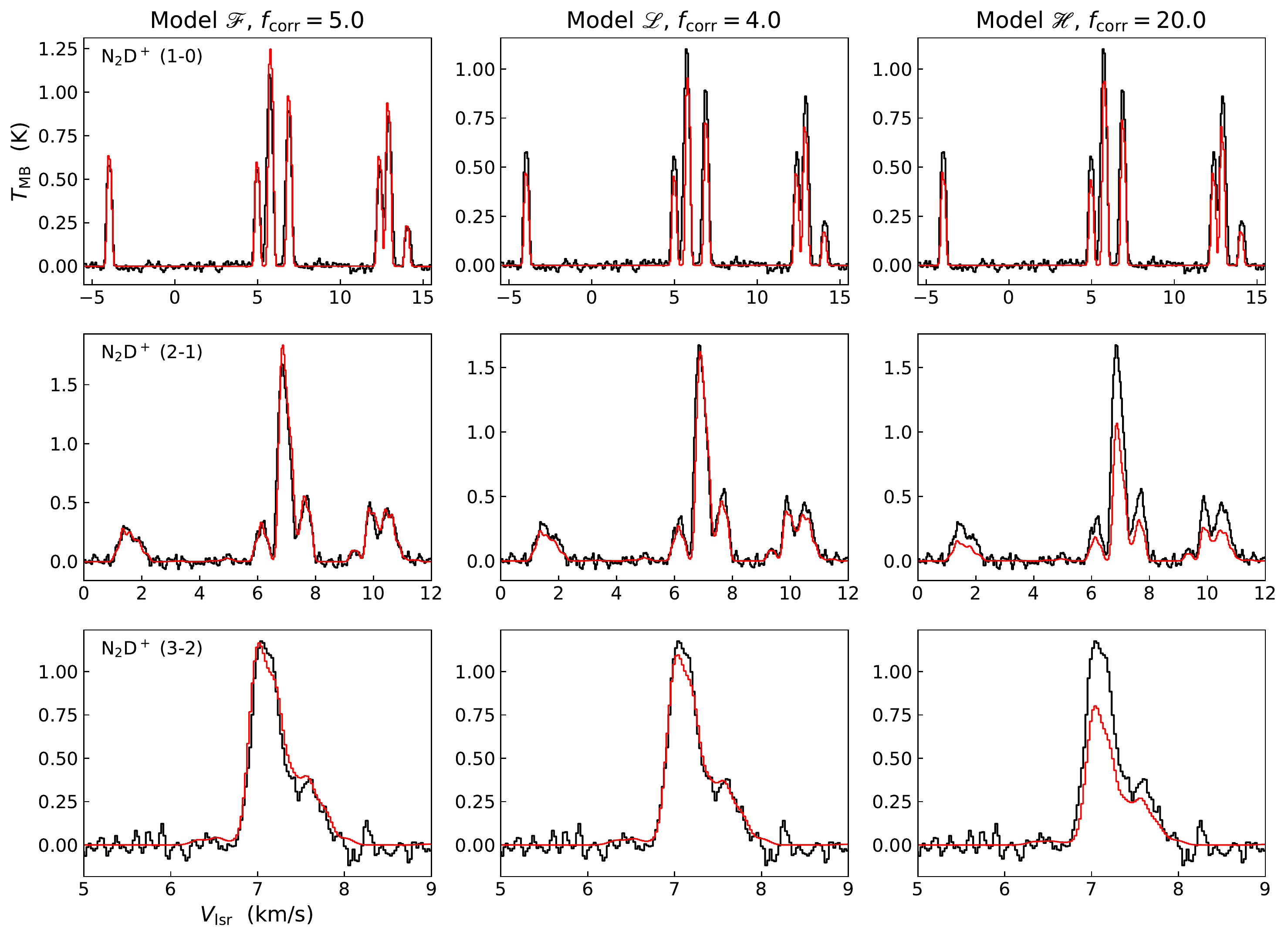}
\caption{As Fig. \ref{Models_n2hp}, but for \nndp. \label{Models_n2dp}}
\end{figure*}

\begin{figure*}[!h]
\centering
\includegraphics[width=.85\textwidth]{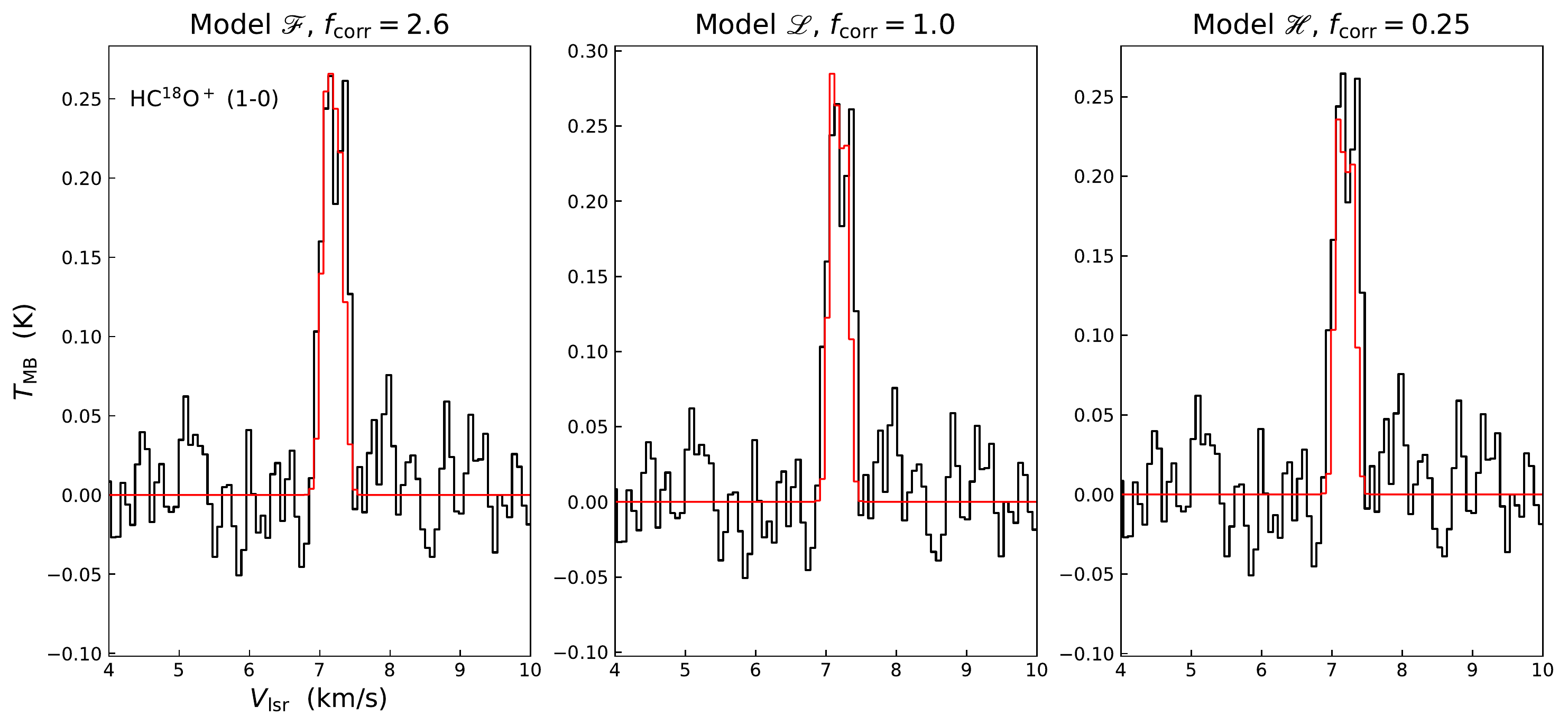}
\caption{As Fig. \ref{Models_n2hp}, but for \hcdop. \label{Models_hcdop}}
\end{figure*}

\begin{figure*}[!h]
\centering
\includegraphics[width=.85\textwidth]{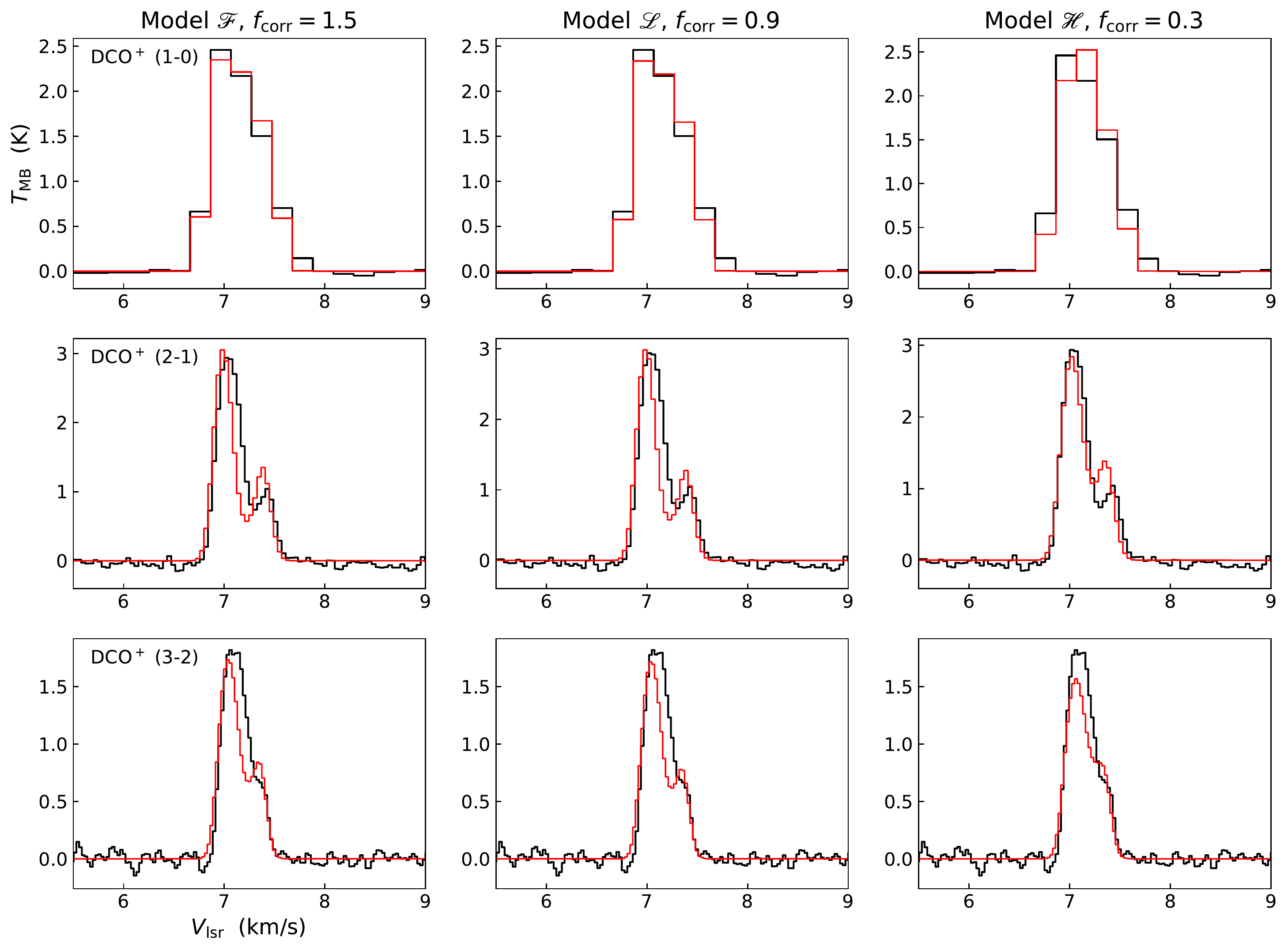}
\caption{As Fig. \ref{Models_n2hp}, but for \dcop. \label{Models_dcop}}
\end{figure*}

In a first test, we focus on the evolutionary time of $ t =  10^6\, \rm yr$, which in RB19 provided the best fit for all the molecules of interest. We hence run MOLLIE with the abundance profiles at this time step, performing a few ($2-5$) tests to find the \xab value that provides a good overall match to the observed fluxes and line shapes for all the observed transitions of each molecule, based in particular on the parameters listed above at the points \ref{list2} and \ref{list3}.  In Appendix \ref{App:xab} we validate this approach to determine \xab through the comparison with a more rigorous, quantitative approach, showing that the two methods produce consistent results within the uncertainties. The best-fit results at $t=10^6 \rm \, yr$ are shown in Figs. \ref{Models_n2hp} to \ref{Models_dcop}, and they are discussed in detail in the following. \par
\paragraph{\nnhp} None of the models at $ t =  10^6\, \rm yr$  is able to reproduce the intensity ratio of the two observed rotational transitions for \nnhp. Models \modf and \modl tend to overestimate the flux of the (3-2) transition with respect to that of the (1-0), whilst the opposite is true for model \modh. This latter, furthermore, requires an enhancement of 400\% of the overall abundance profile in order to match  the observed line intensities. The abundance obtained with model \modf needs to be increased by $\text{\xab} = 1.5$, whilst model \modl predicts an abundance profile that agrees reasonably well with the observations, despite underestimating by 20\% the intensity of the main component of the (1-0) line, and overestimating by 9\% that of the (3-2) line.  
\paragraph{\nndp} All models tend to underestimate the fluxes of the three \nndp transitions, similarly to what was found in RB19. Model \modh is characterised by the strongest underestimation, and the \nndp abundance must be enhanced of a factor of 20 in order to match the observed (1-0) line. Still the higher $J$-transitions appear underestimated. Model \modl requires the smallest \xab factor, and the synthetic line profiles are in good agreement with the observed ones, with the exception of the (1-0) transition, which is underestimated by $\approx 15$\%.

\paragraph{\hcdop} Also for this molecule, model \modl provides the best agreement in terms of flux. None of the models however is able to reproduce the symmetric double-peak features seen in the spectrum, which may depend on the kinematic structure of the source (Ferrer et al., in prep). 
\paragraph{\dcop} This molecule is the one that requires the smallest \xab factors among the targeted species, and is the one for which the intensity ratios of the different $J$-transitions are better reproduced. Model \modh has to be reduced by a factor of $\approx 3$ to reproduce the observed fluxes, whereas model \modf has to be enhanced by $\text{\xab} = 1.5$. For model \modl, a modification of 10\% is enough to obtain a good agreement with the observations. 
\par
Based on these results, we conclude first of all that the model \modh, with the highest \crir profile, is excluded by our data. In particular, it underestimates severely the abundance of \nndp, and to a lesser extent also that of \nnhp, whilst it overestimates the abundances of \hcop isotopologues. Models \modf and  \modl are instead in better agreement with the observations The latter appears better than model \modf, since it requires the smallest \xab corrective factors for three out of the four targeted species. It is in fact remarkable that a single chemical model is able to reproduce simultaneously several transitions of \nnhp, \hcdop, and \dcop (with a correction of only 10\%), at the same time underestimating the fluxes of the \nndp lines by the smallest factor. 
\par
The bottom panels of Fig. \ref{ChemMod} show the ratio between the models (\modl/\modf and \modh/\modf), compared with the inverse ratio of the corresponding \xab values used to obtain the best-fit solutions. Qualitatively, the range of radii where the model ratios are similar to the \xab ratios gives us an idea of which parts of the source contribute the most to the observed transitions. For most of the combinations of models and species, it can be seen that this agreement is found for {radii} smaller than a few $\times 10^{-2} \, \rm pc$, corresponding to volume densities higher than $\approx 10^5 \, \rm cm^{-3}$ (see left panel of Fig. \ref{CRIR}). This is consistent with the high critical densities of the analysed transitions, which are in the range from $\text{a few } 10^4 \, \rm cm^{-3} \text{ to } \approx  10^6 \, \rm cm^{-3} $.

\subsection{Testing different evolutionary stages\label{diff_times}}
The previous subsection has shown that model \modl provides the best fit to the observations at the evolutionary time of one million years. We can however test whether different times produce better agreements. To do so, we have run the non-LTE radiative transfer on model \modl at three further time steps: $ t =  10^5\, \rm yr$,  $ t =  5\times 10^5\, \rm yr$, and $ t =  5 \times10^6\, \rm yr$. These abundance profiles are shown in Appendix \ref{AllModels} (together with the ones for models \modf and \modh). As previously described, we tested a few \xab values in order to match the observed fluxes. \par

There are two main ways in which the changes in abundance as a function of time affect the simulated line profiles. First of all, the chemical evolution will lead to an overall increase or decrease of the total molecular column density, which {modify the} predicted fluxes. Secondly, as time goes on, depletion (due to freeze out onto the dust grains, either of the targeted molecule or of its precursors) becomes more significant. This changes the intensity ratios of the different transitions, since the higher-$J$ transition (having higher $n_\mathrm{c}$) will be less excited with respect to the (1-0) lines. This effect was already observed in RB19, and it stresses again how crucial it is to have access to distinct transitions with different critical density of the same molecular tracer to constrain the molecular abundance profile.
\par
Concerning \nnhp, no time step improves significantly the agreement with the observations. In particular, the model at $t=10^5\,$yr does not match the intensity ratios of the (3-2) to the (1-0) line and of the different hyperfine components in the (1-0) transition. The model at $t=5 \times 10^6\,$yr underestimates the observed peak intensity by more than 50\% for both transitions. The models at $t=5 \times 10^5\,$yr and  $ t =  10^6\, \rm yr$ are similar in terms of the ratio of the two rotational lines intensities, with the model at $ t =  10^6\, \rm yr$ providing a slightly better agreement.\par

\begin{figure}[!h]
\centering
\includegraphics[width=0.5\textwidth]{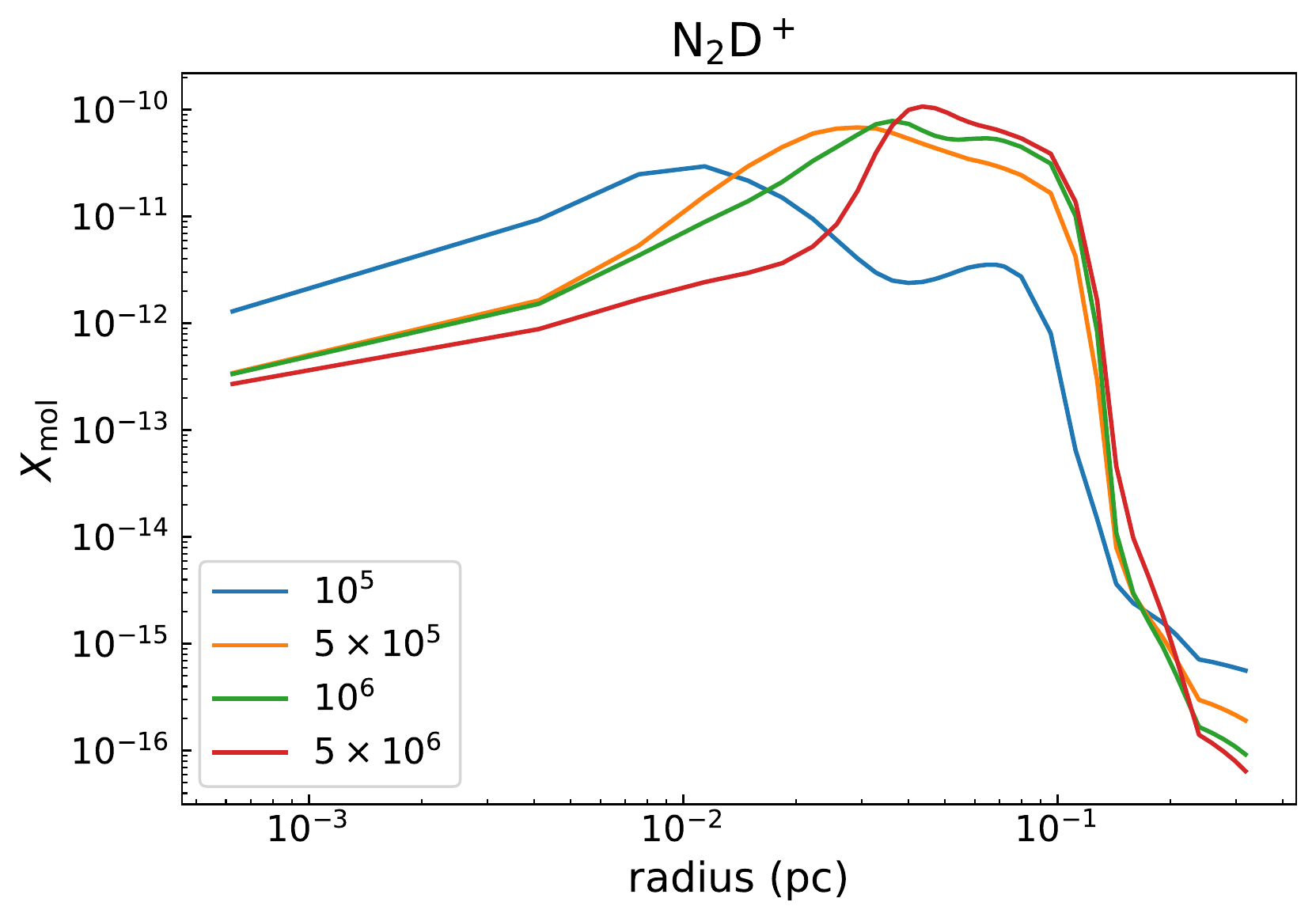}
\caption{Abundance profiles for \nndp in model \modl at four different time steps, from $ t =  10^5\, \rm yr$ to $ t =  5 \times10^6\, \rm yr$. One can see that as the chemistry evolves, \nndp is affected by stronger depletion, with a higher contrast between the central abundance and the peak value. \label{n2dp_modl}}
\end{figure}

The lines of \nndp, on the other hand, are better reproduced when adopting an earlier evolutionary time ($ t =  5 \times10^5\, \rm yr$), and $\text{\xab} =3$. 
 In order to better understand the evolution of the \nndp profile, we show in Fig. \ref{n2dp_modl} the abundance profiles at four time steps for model \modl. As the chemistry evolves, the peak increases in value and it shifts outwards, whilst at the centre the abundance decreases. In Sect. \ref{mod:1e6} we showed how at $ t = 10^6\, \rm yr$ the line intensity ratios are well reproduced, but the abundance profile has to be increased by a factor $\text{\xab} = 4$. The earliest and the latest time steps perform worse because they predict too little or too strong depletion, respectively. In the former case ($ t = 10^5\, \rm yr$) the fluxes of the (2-1) and (3-2) transitions are overestimated with respect to the (1-0) line, and the other way around for the latter case ($ t = 5 \times 10^6\, \rm yr$). The intermediate time step $ t =  5 \times10^5\, \rm yr$ presents instead the best combination of total molecular density and depletion level.

\par
For \dcop and \hcdop, the model at $ t =  10^6\, \rm yr$ requires the smallest corrective factors to reproduce the observed lines, and is hence considered the best-fit model. We highlight that using distinct time steps to reproduce different molecular tracers does not represent a contradiction, due to the fact that our physical model is static, and the chemistry is evolved in post-processing. {The physical age of the source is linked to the dynamical timescale, and not to the chemical one. Moreover, distinct molecules are sensitive to different physical conditions. Species such as \nndp, which is formed only when CO is depleted from the gas phase and temperatures are low, trace mainly the very dense, inner parts of the source, where the dynamical timescale is shorter (since it depends inversely on square root of the volume density). This could explain why \nndp is better reproduced at an earlier chemical timescale with respect to the other species.}

\par
A better performance of the two remaining \crir models can be excluded based on the following arguments. The main problems of model \modh are the severe underestimation of \nndp abundance (by more than one order of magnitude) and a large overestimation of \hcdop abundance (by a factor of $4$). At any time step, model \modh presents the highest abundance of \hcop (and thus of \hcdop), which will always be overestimated. Concerning \nndp, model \modh always predicts the lowest abundance, with the exception of the earliest evolutionary time ($ t = 10^5\, \rm yr$, see Fig. \ref{ChemMod_all}). This time step, however presents almost no signs of molecular depletion, and we have shown how this kind of models are not able to reproduce the intensity ratios of the distinct $J$ transitions. 
\par
The main problems of model \modf at $ t = 10^6\, \rm yr$ are the underestimation of \hcdop (by a factor of 2.6) and of \nndp (by a factor of 5). The former problem will not be solved at any time step, since the \hcop abundance is always the lowest in the fiducial model with respect to the other models. Concerning \nndp, we tested the same time-step that provides the best fit for the model \modl: $ t = 5 \times 10^5\, \rm yr$ and $\text{\xab} = 3$. We find that with t his abundance profile the (3-2) line is reproduced, but the (2-1) and the (1-0) transition are overestimated by $\approx 30$\%. 
We conclude that changing the evolutionary stage does not provide better agreement with the observations for both model \modh and \modf, with respect to model \modl.

\subsection{The electron fraction}
\begin{figure}[!t]
\centering
\includegraphics[width=0.5\textwidth]{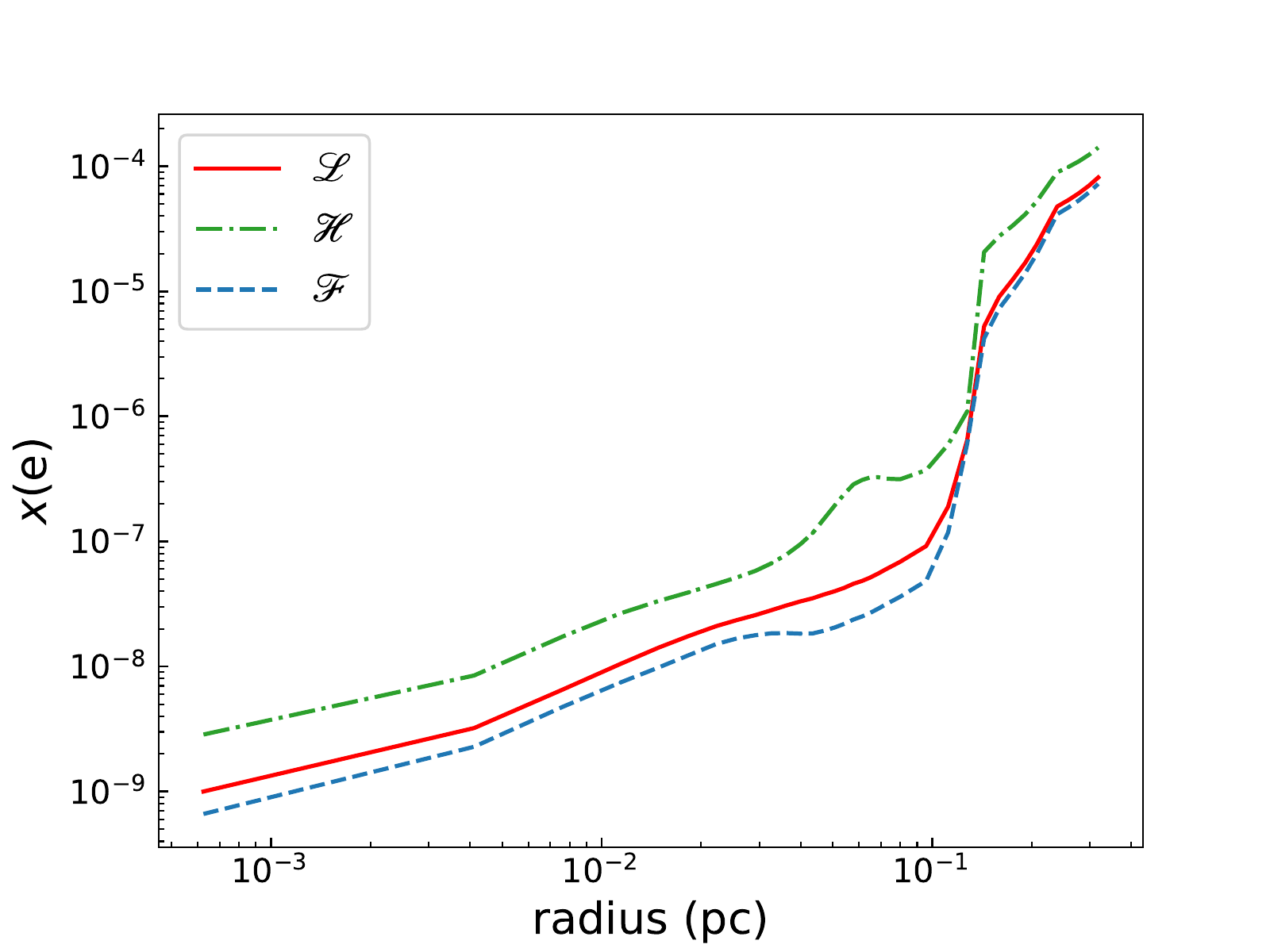}
\caption{ $x(\mathrm{e})$ profiles in model \modf (dashed blue curve), \modl (solid red), and \modh (dashed-dotted {green}). All the models are shown at $t=10^6\rm \, yr$. \label{el_fract}}
\end{figure}

Since they are the main ionising source in dense gas, CRs directly determine the abundance of free electrons, or electron fraction ($x(\mathrm{e})= n(\mathrm{e})/n$). As explained in Sect. \ref{Intro}, this parameter regulates the degree of coupling between the gas and the magnetic fields, and it is linked to the timescale for ambipolar diffusion. The results of the chemical modelling include also this parameter, which is shown in Fig. \ref{el_fract} for the three models \modf, \modl, and \modh. We focus on the model results at $t=10^6\rm \, yr$, but we highlight that in the inner parts of the core ($r<0.05\,$pc), $x\mathrm{(e)}$ values change only by a few percent when varying the time step. \par
Figure \ref{el_fract} shows that the electron fraction increases with increasing \crir, as expected due to the nature of the ionisations. At the core centre, the value obtained in model \modl is $x(\mathrm{e})= 10^{-9}$. This result is in agreement with the one from \cite{Caselli02}, who used more simplified chemical models on a subset of the molecular transitions presented in this work to estimate this quantity. 

\subsection{Simulating the cloud evolution\label{Sec:CloudEvo}}
As a further test on the fiducial model with constant $\text{\crir} = 1.3 \times 10^{-17} \rm \, s^{-1}$, we changed the initial conditions of the chemical model. We simulate first a cloud-like evolution for $t_\mathrm{CE} = 10^6 \, \rm yr$. For this initial simulation, we use the following physical conditions: $n = 10^3 \,  \rm cm^{-3}$, $T = 15 \rm \, K$, and external extinction $ A \rm _V = 2 \,mag$. The final abundances of this model are then taken as initial condition for the simulation of the core. The time step values $t$ refer to the time elapsed after the end of the initial cloud simulation (i.e. we reset $t=0$ at the time $t_\mathrm{CE}$) . \par
\begin{figure*}[!h]
\centering
\includegraphics[width=.8\textwidth]{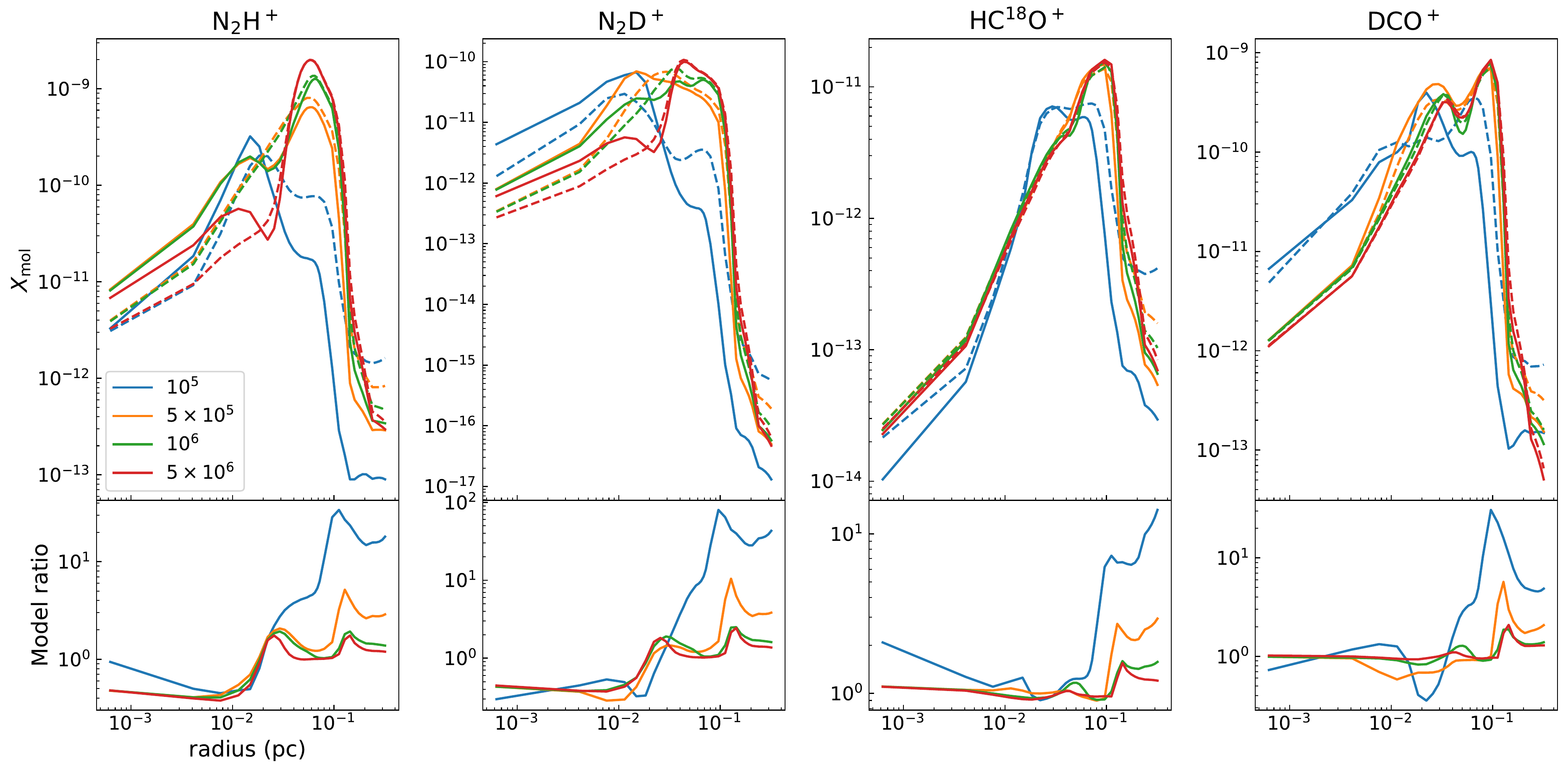}
\caption{\textit{Top panels:} Abundance profiles obtained with the model with constant $\text{\crir} = 1.3 \times 10^{-17} \rm \, s^{-1}$, with no initial cloud evolution (model \modf, solid curves) and with a cloud evolution of $t_\mathrm{CE} = 10^6 \, \rm yr$ to set the initial conditions (dashed curves). The different colours refer to the four time steps investigated in this work (labelled in the first panel, in year). \textit{Bottom panels:}  Ratios between the abundance profiles of the model with and without cloud evolution. The colour code is the same as for the upper panels. \label{CloudEvo}}
\end{figure*}

In Fig. \ref{CloudEvo} we show the final abundance profiles for the species of interest from the two models, with and without cloud evolution. \nnhp and \nndp show the strongest variations. The model with cloud evolution presents stronger depletion towards the centre at any time step. For \hcop isotopologues, the differences are smaller, especially for \hcop. In the case of \dcop, the molecular abundance is generally smaller in the model with cloud evolution, with the exception of the earliest time step ($ t =  10^5\, \rm yr$). \par
\begin{figure*}[!h]
\centering
\includegraphics[width=.8\textwidth]{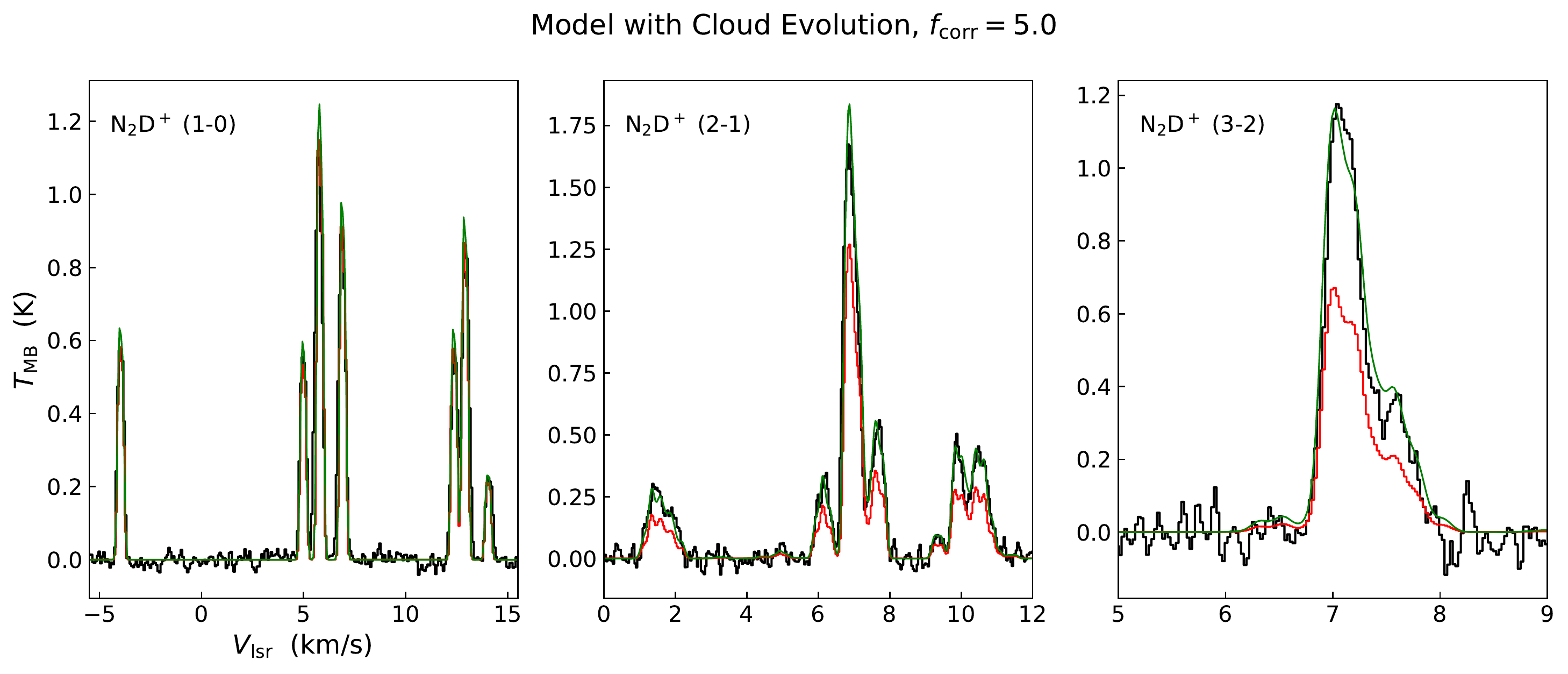}
\caption{Red curves show the best-fit model for \nndp lines obtained with the fiducial model with initial cloud evolution and $\zeta_2 = 1.3 \times 10^{-17} \, \rm s^{-1}$, at $ t =  10^6\, \rm yr$, overlaid to the observations (black). The used \xab value is indicated at the top.  The modelled lines produced using model \modf with no cloud evolution at $ t =10^6\, \rm yr$ (and $\text{\xab} = 5)$ are shown in green. It is clear that the model with cloud evolution underestimates the (2-1) and (3-2) lines. \label{n2dp_modCE}}
\end{figure*}

As a result of these differences, the cloud-evolution model requires in general higher corrective factor \xab to reproduce the observed fluxes. We report in Appendix \ref{App:cloud_evo} the complete set of comparisons between the synthetic and observed spectra at $t = 10^6 \, \rm yr$. We discuss here the case of \nndp, which, as shown also in the previous subsections, is sensitive to the level of depletion due to the high critical densities of the (2-1) and (3-2) rotational lines. We report the results of MOLLIE for this model, multiplied by $\text{\xab} = 5$, in Fig. \ref{n2dp_modCE}, compared with the best-fit solution found for model \modl. Whilst the first rotational transition is well reproduced both in its line profile and in the intensity ratios of the distinct hyperfine components, the (2-1) and (3-2) transitions are underestimated by 25\% and 45\% of the peak intensity, respectively. These results for \nndp, together with the fact that the corrective factors \xab are higher also for the other species (see Appendix \ref{App:cloud_evo}), suggest that the initial conditions obtained with the initial cloud evolution model are not consistent with our observations. \par
We point out that there are significant uncertainties involved in modelling an initial cloud-like evolution, associated for instance with choosing the appropriate physical conditions for the cloud stage, or selecting the value of $t_\mathrm{CE} $. Furthermore, neglecting the dynamical evolution from cloud to core is also a source of uncertainty. 

\subsection{Constant $\text{\crir} = 3 \times 10^{-17} \rm \, s^{-1}$ model}
The models so far explored are characterised by average \crir values that are significantly different:  $\text{\crir} = \text{const} =  1.3 \times 10^{-17} \rm \, s^{-1}$ in model \modf,  $\langle \text{\crir} \rangle = 3 \times 10^{-17} \rm \, s^{-1}$ in model \modl, and  $\langle \text{\crir} \rangle = 2 \times 10^{-16} \rm \, s^{-1}$ in model \modh. The last two values refer to gas column densities in the range $N =[0.2-5]\times 10^{22} \, \rm cm^{-2}$ (see Fig. \ref{CRIR}), obtained by integrating the volume density profile from \cite{Keto15}.  It is hence important to verify whether our observations are truly sensitive to the CR attenuation towards the core centre, or if they can only distinguish among different average values of \crir. \par

\begin{figure*}[!h]
\centering
\includegraphics[width=.85\textwidth]{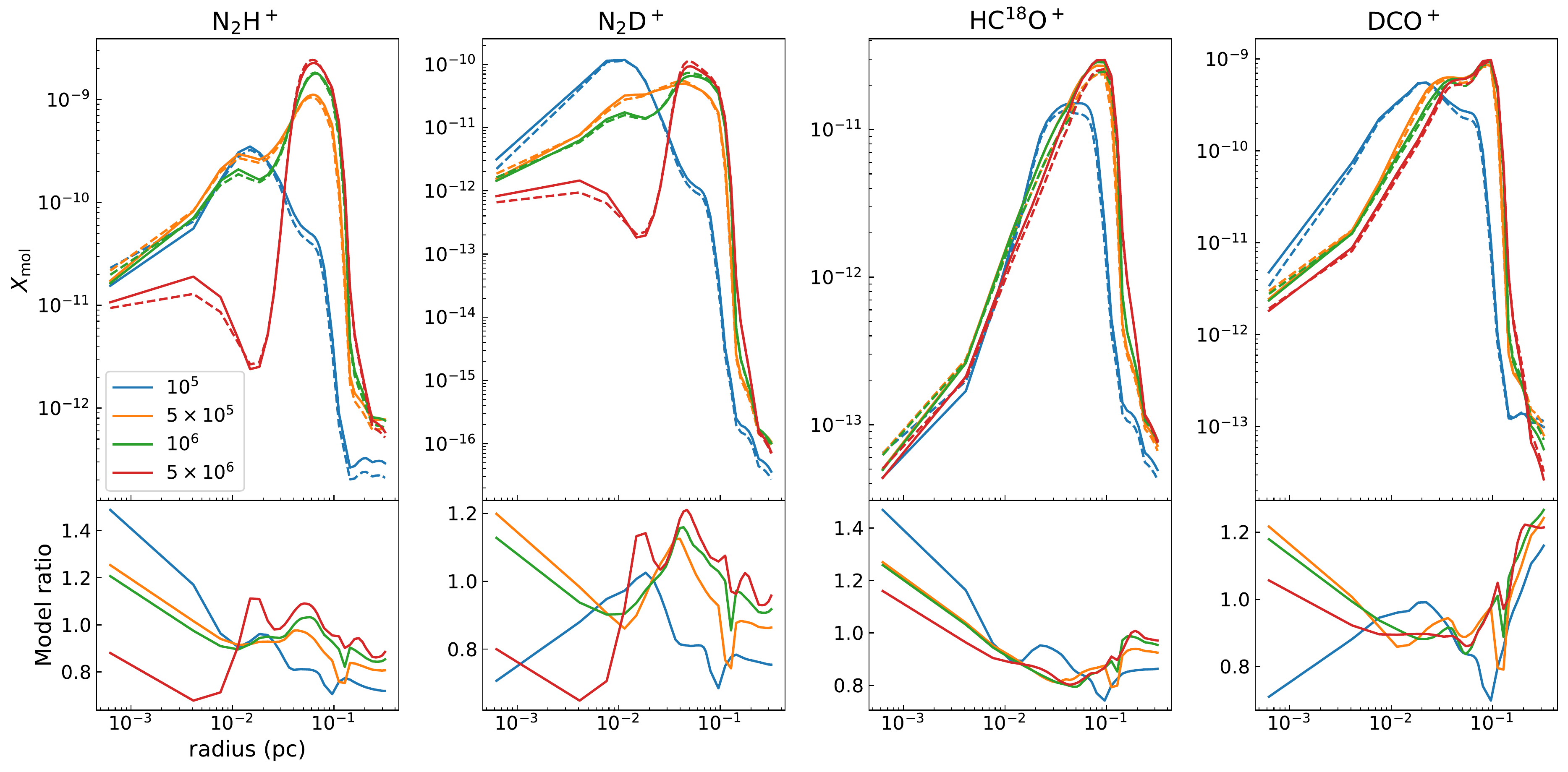}
\caption{\textit{Top panels:} Abundance profiles obtained from the model \modl (solid lines), and from the model with constant $\text{\crir} = 3 \times 10^{-17} \rm \, s^{-1}$ (dashed lines). The different colours refer to the four time steps investigated in this work (labelled in the first panel, in year). \textit{Bottom panels:} Ratios between the abundance profiles from model \modl and from the model with constant $\text{\crir} = 3 \times 10^{-17} \rm \, s^{-1}$. The colour code is the same as for the upper panels. \label{EnhancedMod}}
\end{figure*}

In order to test this, we have run another chemical model, identical to the fiducial one but adopting a constant CR ionisation rate equal to $\text{\crir} = 3 \times 10^{-17} \rm \, s^{-1}$ (the average value of model \modl). The resulting abundance profiles are shown in Fig. \ref{EnhancedMod}, in comparison to those of the model \modl. The two models produce very similar profiles, in particular at the two time steps that provide the best-fit ($ t = 5 \times 10^5\, \rm yr$ and $ t =  10^6\, \rm yr$). The median  relative differences at these time steps are $6-9$\% for \nnhp, $8-10$\% for \nndp, $16-17$\% for \hcop, and $10-13$\% for \dcop. As a consequence, also the synthetic line profiles, shown in {Appendix} \ref{App:enhanchedMod}, appear similar.
\par
Given the uncertainties of this work (e.g. the assumed physical model, or the uncertainties relative to the chemical network, such as the rate coefficients), we cannot make stringent conclusions on these two models. In particular, we lack the spatial resolution to conclusively assess if the molecular line data support the model of CR attenuation in L1544. We highlight that, in fact, the angular resolution of the low-$J$ transitions are in the $27-36''$ range, which correspond to $0.018-0.024\, \rm pc$  ($3600-4900\, \rm AU$) at the distance of L1544.

\par
\cite{Neufeld17} used observations of several ions (and in particular of $\rm H_3^+$) to derive the CR ionisation rate in a sample of diffuse clouds. Their data supported a stronger attenuation of the CRs with increasing density with respect to the models of PI18, with a power-law dependence following $\zeta_2 \propto N ^{-0.92}$ (see their Figure 6). The reader may note that in the range of column densities spanned by the physical model, the model \modl predicts a shallow dependency on the column density, approximately following $\zeta_2 \propto N ^{-0.2}$. We tested the steeper relation derived from \cite{Neufeld17} in our chemical model\footnote{The relation derived from \cite{Neufeld17} reads $
\log_{10}(\zeta_2/\mathrm{s^{-1}}) = -0.92\times \log_{10}(N/\mathrm{cm}^{-2})+3.9$.}. The resulting abundances underestimate the observed transitions with respect to model \modl. The reason for this behaviour is that the relation found by \cite{Neufeld17}, when applied to the range of densities of L1544, implies that for radii smaller than $0.01\, \rm pc$ the CR ionisation rate is $\zeta_2 < 3 \times 10^{-17} \rm \, s^{-1}$. As a consequence, in the central regions ---where the targeted transitions are excited, see Sect. \ref{mod:1e6}--- the lower CR ionisation rate leads in general to {a} lower predicted abundance, in particular for \hcdop and \dcop, for which the dependence on \crir is monotonic, as discussed in Sect. \ref{Results}.

\section{Summary and Conclusions \label{Concl}}
In this work we have studied the effect of the CR ionisation rate on the results of gas-grain chemical simulations, focusing on four species: \nnhp, \nndp, \hcdop, and \dcop. The predicted abundance profiles have been used to perform a full non-LTE radiative transfer analysis, and the synthetic spectra have been compared with recent, high-sensitivity spectra of the molecules at the dust peak of the prototypical pre-stellar core L1544.
\par
The first important conclusion is that the model with high \crir (\modh from PI18), characterised by an average value of $\langle \text{\crir} \rangle = 2 \times 10^{-16} \rm \, s^{-1}$, is excluded by our observations. In particular, this model strongly underestimates the \nndp abundance, while it overestimates by a factor of $3-4$ the intensities of \hcdop and \dcop lines. 
 \par
The model \modl from PI18 is the one that provides the overall best agreement with the observations. It is remarkable that, with the exception of \nndp ---which is underestimated by all the models here considered--- the other targeted molecules are reproduced within 10\%. The fiducial model, with standard $\zeta_2 = 1.3 \times 10^{-17} \rm \, s^{-1}$, performs worse than model \modl, and requires corrections in the range $\text{\xab}=1.5-5$ at $t=  10^{6} \, \rm yr$ to reproduce the observed rotational lines. Changing the initial conditions by simulating first a cloud-like evolution of the chemistry for  $t_\mathrm{CE}=  10^{6} \, \rm yr$ does not improve the agreement. Table \ref{xab_summ} contains a summary of the \xab factors used in the different models. We highlight that it is not a contradiction to use different evolutionary stages for the distinct molecules, since the chemical model is run in post-processing with respect to the dynamical evolution. The fact that \nndp requires a shorter $t$ with respect to the other species may be due to the fact that this deuterated species, requiring significant CO freeze-out for its efficient production, traces mainly the central parts of the pre-stellar core, where we expect the dynamical time to be the shortest due to the high volume densities. 
\begin{table*}[!h]
\renewcommand{\arraystretch}{1.4}
\centering
\caption{Summary of the \xab values used in the different chemical models for each molecular species.\label{xab_summ} }
\begin{tabular}{c|cccc}
\hline
Model                                                      & \multicolumn{4}{c}{\xab}       \\
                                                           & \nnhp & \nndp & \hcdop & \dcop \\
  \hline \hline                                                         
\modf                                                      & 1.5   & 5.0   & 2.6    & 1.5   \\
\modl                                                      & 1.0   & 3.0\tablefootmark{a}   & 1.0    & 0.9   \\
\modh                                                      & 4.0   & 20.0  & 0.25   & 0.3   \\
\modf with cloud evolution                                     & 2.0   & 5.0   & 2.5    & 1.6   \\
$\zeta_2 = \text{const} = 3 \times 10^{-17} \, \rm s^{-1}$ & 1.0   & 3.0\tablefootmark{a}   & 1.0    & 0.9  \\
\hline
\end{tabular}
\tablefoot{
\tablefoottext{a}{Models taken at $t = 5 \times 10^{5} \, \rm yr$ }. All the other models are taken at $t =  10^{6} \, \rm yr$.}
\end{table*}
\par
In order to check if our observations are sensitive to the attenuation of \crir, or more to its average value across the core, we have run another chemical model using $\zeta_2 = 3 \times 10^{-17} \rm \, s^{-1}$ (the average \crir of model \modl at column densities of $N = [0.2-5]\times 10^{22} \, \rm cm^{-2}$). The estimated abundance profiles are very similar to model \modl, especially at radii $r > 0.05 \rm \, pc = 10^4 \, AU $. As a consequence, the synthetic spectra are also similar, and do not allow us to distinguish between the cases of constant vs. attenuated \crir. We tested also the attenuation model from \cite{Neufeld17}, which predicts a steeper decrease of \crir with increasing density ($\zeta_2 \propto N^{-0.92}$). The latter is in worse agreement with the observations with respect to model \modl, since it predicts CR ionisation rate values lower than model \modl in the innermost parts of the core ($< 0.01\, \rm pc = 2000 \, AU$), from where the targeted transitions mostly arise. In order to constrain the decrease of \crir with increasing density, two complementary approaches appear necessary: \textit{i)} performing observations at higher spatial resolution (hundreds of astronomical units) of molecular lines with high critical densities, such as the one used for this work; \textit{ii)} targeting transitions with lower critical densities ($10^3-10^4 \, \rm cm^{-3}$), which could trace the \crir value in the low-density extended envelope.
\par
Further improvements on this topic could come from expanding the sample by repeating the analysis on other prestellar/protostellar cores. In this respect, we highlight that the two species that present the strongest sensitivity in their simulated spectra to changing \crir are \hcdop and \nndp. The peak intensity of the \hcdop (1-0) transition, in fact, changes by a factor of $\approx 10$ going from model \modf to \modh. \nndp is also sensitive to high \crir values, which makes the abundance of this molecule drop (see for instance model \modh). This, combined with the fact that \nndp has three transitions easily accessible at millimetre wavelengths, and with high critical densities, makes this species a good probe of the CR ionisation rate in dense cores.
\par
Our results appear to be in agreement with those of \cite{Bovino20}, who adopted an analytical approach and obtained $\zeta_2 = 2-3\times 10^{-17} \rm \, s^{-1} $ in L1544, and also with those of \cite{VanDerTak00}, obtained in a different molecular cloud. {Moreover, they are consistent with \cite{Maret07}, who estimated $\zeta_2 = [1.5-9]\times 10^{-17} \, \rm s^{-1}$ in the starless core Barnard 68, and within uncertainties with \cite{Maret13}, who found that $\zeta_2 \approx 4.6\times 10^{-17} \rm \, s^{-1} $ reproduces reasonably the $\rm C^{18}O$ and $\rm H^{13}CO^+$ emission in two pre-stellar cores}. It is worth noticing that in the recent work from  \cite{Ivlev19}, the authors estimated an upper limit of $\text{\crir} \approx 10^{-16} \rm \, s^{-1} $ using the gas temperature measurements in L1544. This upper limit, which is still consistent with our findings, was derived assuming unevolved dust with the MRN size distribution \citep{Mathis77}; if small grains are eliminated due to a rapid coagulation, which is expected to occur in molecular clouds \citep{Silsbee20}, the upper limit on \crir is a factor of $\approx 3$ lower.  \par
One of the {limitations} of this work is that we are not considering the effects of magnetic fields. As charged particles, CR are affected by magnetic forces, and they gyrate around the magnetic field lines during their propagation. Hence, depending on the magnetic field morphology, and in particular on the ratio between the toroidal and the poloidal components of the magnetic field, CRs experience an effective column density which can be much larger than the observed one. In particular at the core centre, where the field lines are expected to be strongly entangled, the effective column density ``seen'' by CRs can be significantly higher than $N$, causing a strong decrease of \crir \citep[see for instance][]{Padovani13}. Future polarisation observations aimed at recovering the morphology of the magnetic field in L1544 will provide useful constraints in this sense.


%

\begin{thebibliography}{50}
\expandafter\ifx\csname natexlab\endcsname\relax\def\natexlab#1{#1}\fi

\bibitem[{{Bizzocchi} {et~al.}(2013){Bizzocchi}, {Caselli}, {Leonardo}, \&
  {Dore}}]{Bizzocchi13}
{Bizzocchi}, L., {Caselli}, P., {Leonardo}, E., \& {Dore}, L. 2013, \aap, 555,
  A109

\bibitem[{{Black} {et~al.}(1978){Black}, {Hartquist}, \& {Dalgarno}}]{Black78}
{Black}, J.~H., {Hartquist}, T.~W., \& {Dalgarno}, A. 1978, \apj, 224, 448

\bibitem[{Bovino {et~al.}(2020)Bovino, Ferrada-Chamorro, Lupi, Schleicher, \&
  Caselli}]{Bovino20}
Bovino, S., Ferrada-Chamorro, S., Lupi, A., Schleicher, D. R.~G., \& Caselli,
  P. 2020, MNRAS: Letters, 495, L7

\bibitem[{{Caselli} {et~al.}(2019){Caselli}, {Pineda}, {Zhao}, {Walmsley},
  {Keto}, {Tafalla}, {Chac{\'o}n-Tanarro}, {Bourke}, {Friesen}, {Galli}, \&
  {Padovani}}]{Caselli19}
{Caselli}, P., {Pineda}, J.~E., {Zhao}, B., {et~al.} 2019, \apj, 874, 89

\bibitem[{{Caselli} {et~al.}(1998){Caselli}, {Walmsley}, {Terzieva}, \&
  {Herbst}}]{Caselli98}
{Caselli}, P., {Walmsley}, C.~M., {Terzieva}, R., \& {Herbst}, E. 1998, \apj,
  499, 234

\bibitem[{{Caselli} {et~al.}(2002){Caselli}, {Walmsley}, {Zucconi}, {Tafalla},
  {Dore}, \& {Myers}}]{Caselli02}
{Caselli}, P., {Walmsley}, C.~M., {Zucconi}, A., {et~al.} 2002, \apj, 565, 344

\bibitem[{{Chac{\'o}n-Tanarro} {et~al.}(2017){Chac{\'o}n-Tanarro}, {Caselli},
  {Bizzocchi}, {Pineda}, {Harju}, {Spaans}, \& {D{\'e}sert}}]{ChaconTanarro17}
{Chac{\'o}n-Tanarro}, A., {Caselli}, P., {Bizzocchi}, L., {et~al.} 2017, \aap,
  606, A142

\bibitem[{{Crapsi} {et~al.}(2007){Crapsi}, {Caselli}, {Walmsley}, \&
  {Tafalla}}]{Crapsi07}
{Crapsi}, A., {Caselli}, P., {Walmsley}, M.~C., \& {Tafalla}, M. 2007, \aap,
  470, 221

\bibitem[{{Cummings} {et~al.}(2016){Cummings}, {Stone}, {Heikkila}, {Lal},
  {Webber}, {J{\'o}hannesson}, {Moskalenko}, {Orlando}, \&
  {Porter}}]{Cummings16}
{Cummings}, A.~C., {Stone}, E.~C., {Heikkila}, B.~C., {et~al.} 2016, \apj, 831,
  18

\bibitem[{{Dalgarno} \& {Lepp}(1984)}]{Dalgarno84}
{Dalgarno}, A. \& {Lepp}, S. 1984, \apj, 287, L47

\bibitem[{{Doty} {et~al.}(2004){Doty}, {Sch{\"o}ier}, \& {van
  Dishoeck}}]{Doty04}
{Doty}, S.~D., {Sch{\"o}ier}, F.~L., \& {van Dishoeck}, E.~F. 2004, \aap, 418,
  1021

\bibitem[{{Glassgold} \& {Langer}(1974)}]{Glassgold74}
{Glassgold}, A.~E. \& {Langer}, W.~D. 1974, \apj, 193, 73

\bibitem[{{Indriolo} {et~al.}(2007){Indriolo}, {Geballe}, {Oka}, \&
  {McCall}}]{Indriolo07}
{Indriolo}, N., {Geballe}, T.~R., {Oka}, T., \& {McCall}, B.~J. 2007, \apj,
  671, 1736

\bibitem[{{Indriolo} \& {McCall}(2012)}]{Indriolo12}
{Indriolo}, N. \& {McCall}, B.~J. 2012, \apj, 745, 91

\bibitem[{{Ivlev} {et~al.}(2015){Ivlev}, {Padovani}, {Galli}, \&
  {Caselli}}]{Ivlev15}
{Ivlev}, A.~V., {Padovani}, M., {Galli}, D., \& {Caselli}, P. 2015, \apj, 812,
  135

\bibitem[{{Ivlev} {et~al.}(2021){Ivlev}, {Silsbee}, {Padovani}, \&
  {Galli}}]{Ivlev21}
{Ivlev}, A.~V., {Silsbee}, K., {Padovani}, M., \& {Galli}, D. 2021, \apj, 909,
  107

\bibitem[{{Ivlev} {et~al.}(2019){Ivlev}, {Silsbee}, {Sipil{\"a}}, \&
  {Caselli}}]{Ivlev19}
{Ivlev}, A.~V., {Silsbee}, K., {Sipil{\"a}}, O., \& {Caselli}, P. 2019, \apj,
  884, 176

\bibitem[{{Keto} {et~al.}(2015){Keto}, {Caselli}, \& {Rawlings}}]{Keto15}
{Keto}, E., {Caselli}, P., \& {Rawlings}, J. 2015, \mnras, 446, 3731

\bibitem[{{Keto} {et~al.}(2004){Keto}, {Rybicki}, {Bergin}, \&
  {Plume}}]{Keto04}
{Keto}, E., {Rybicki}, G.~B., {Bergin}, E.~A., \& {Plume}, R. 2004, \apj, 613,
  355

\bibitem[{{Keto}(1990)}]{Keto90}
{Keto}, E.~R. 1990, \apj, 355, 190

\bibitem[{{Kim} {et~al.}(2000){Kim}, {Santos}, \& {Parente}}]{Kim00}
{Kim}, Y.-K., {Santos}, J.~P., \& {Parente}, F. 2000, \pra, 62, 052710

\bibitem[{{Maret} \& {Bergin}(2007)}]{Maret07}
{Maret}, S. \& {Bergin}, E.~A. 2007, \apj, 664, 956

\bibitem[{{Maret} {et~al.}(2013){Maret}, {Bergin}, \& {Tafalla}}]{Maret13}
{Maret}, S., {Bergin}, E.~A., \& {Tafalla}, M. 2013, \aap, 559, A53

\bibitem[{{Mathis} {et~al.}(1977){Mathis}, {Rumpl}, \& {Nordsieck}}]{Mathis77}
{Mathis}, J.~S., {Rumpl}, W., \& {Nordsieck}, K.~H. 1977, \apj, 217, 425

\bibitem[{{Mouschovias} \& {Spitzer}(1976)}]{Mouschovias76}
{Mouschovias}, T.~C. \& {Spitzer}, L., J. 1976, \apj, 210, 326

\bibitem[{{Neufeld} \& {Wolfire}(2017)}]{Neufeld17}
{Neufeld}, D.~A. \& {Wolfire}, M.~G. 2017, \apj, 845, 163

\bibitem[{{Padovani} {et~al.}(2009){Padovani}, {Galli}, \&
  {Glassgold}}]{Padovani09}
{Padovani}, M., {Galli}, D., \& {Glassgold}, A.~E. 2009, \aap, 501, 619

\bibitem[{{Padovani} {et~al.}(2013){Padovani}, {Hennebelle}, \&
  {Galli}}]{Padovani13}
{Padovani}, M., {Hennebelle}, P., \& {Galli}, D. 2013, \aap, 560, A114

\bibitem[{{Padovani} {et~al.}(2018){Padovani}, {Ivlev}, {Galli}, \&
  {Caselli}}]{Padovani18}
{Padovani}, M., {Ivlev}, A.~V., {Galli}, D., \& {Caselli}, P. 2018, \aap, 614,
  A111

\bibitem[{{Phan} {et~al.}(2021){Phan}, {Schulze}, {Mertsch}, {Recchia}, \&
  {Gabici}}]{Phan21}
{Phan}, V. H.~M., {Schulze}, F., {Mertsch}, P., {Recchia}, S., \& {Gabici}, S.
  2021, arXiv e-prints, arXiv:2105.00311

\bibitem[{{Planck Collaboration} {et~al.}(2016){Planck Collaboration}, {Ade},
  {Aghanim}, {Alves}, {Arnaud}, {Arzoumanian}, {Ashdown}, {Aumont},
  {Baccigalupi}, {Band ay}, {Barreiro}, {Bartolo}, {Battaner}, {Benabed},
  {Beno{\^\i}t}, {Benoit-L{\'e}vy}, {Bernard}, {Bersanelli}, {Bielewicz},
  {Bock}, {Bonavera}, {Bond}, {Borrill}, {Bouchet}, {Boulanger}, {Bracco},
  {Burigana}, {Calabrese}, {Cardoso}, {Catalano}, {Chiang}, {Christensen},
  {Colombo}, {Combet}, {Couchot}, {Crill}, {Curto}, {Cuttaia}, {Danese},
  {Davies}, {Davis}, {de Bernardis}, {de Rosa}, {de Zotti}, {Delabrouille},
  {Dickinson}, {Diego}, {Dole}, {Donzelli}, {Dor{\'e}}, {Douspis}, {Ducout},
  {Dupac}, {Efstathiou}, {Elsner}, {En{\ss}lin}, {Eriksen},
  {Falceta-Gon{\c{c}}alves}, {Falgarone}, {Ferri{\`e}re}, {Finelli}, {Forni},
  {Frailis}, {Fraisse}, {Franceschi}, {Frejsel}, {Galeotta}, {Galli}, {Ganga},
  {Ghosh}, {Giard}, {Gjerl{\o}w}, {Gonz{\'a}lez-Nuevo}, {G{\'o}rski},
  {Gregorio}, {Gruppuso}, {Gudmundsson}, {Guillet}, {Harrison}, {Helou},
  {Hennebelle}, {Henrot-Versill{\'e}}, {Hern{\'a}ndez-Monteagudo}, {Herranz},
  {Hildebrand t}, {Hivon}, {Holmes}, {Hornstrup}, {Huffenberger}, {Hurier},
  {Jaffe}, {Jaffe}, {Jones}, {Juvela}, {Keih{\"a}nen}, {Keskitalo}, {Kisner},
  {Knoche}, {Kunz}, {Kurki-Suonio}, {Lagache}, {Lamarre}, {Lasenby},
  {Lattanzi}, {Lawrence}, {Leonardi}, {Levrier}, {Liguori}, {Lilje},
  {Linden-V{\o}rnle}, {L{\'o}pez-Caniego}, {Lubin}, {Mac{\'\i}as-P{\'e}rez},
  {Maino}, {Mandolesi}, {Mangilli}, {Maris}, {Martin},
  {Mart{\'\i}nez-Gonz{\'a}lez}, {Masi}, {Matarrese}, {Melchiorri}, {Mendes},
  {Mennella}, {Migliaccio}, {Miville-Desch{\^e}nes}, {Moneti}, {Montier},
  {Morgante}, {Mortlock}, {Munshi}, {Murphy}, {Naselsky}, {Nati},
  {Netterfield}, {Noviello}, {Novikov}, {Novikov}, {Oppermann}, {Oxborrow},
  {Pagano}, {Pajot}, {Paladini}, {Paoletti}, {Pasian}, {Perotto}, {Pettorino},
  {Piacentini}, {Piat}, {Pierpaoli}, {Pietrobon}, {Plaszczynski},
  {Pointecouteau}, {Polenta}, {Ponthieu}, {Pratt}, {Prunet}, {Puget}, {Rachen},
  {Reinecke}, {Remazeilles}, {Renault}, {Renzi}, {Ristorcelli}, {Rocha},
  {Rossetti}, {Roudier}, {Rubi{\~n}o-Mart{\'\i}n}, {Rusholme}, {Sandri},
  {Santos}, {Savelainen}, {Savini}, {Scott}, {Soler}, {Stolyarov}, {Sudiwala},
  {Sutton}, {Suur-Uski}, {Sygnet}, {Tauber}, {Terenzi}, {Toffolatti}, {Tomasi},
  {Tristram}, {Tucci}, {Umana}, {Valenziano}, {Valiviita}, {Van Tent},
  {Vielva}, {Villa}, {Wade}, {Wandelt}, {Wehus}, {Ysard}, {Yvon}, \&
  {Zonca}}]{PlanckXXXV}
{Planck Collaboration}, {Ade}, P.~A.~R., {Aghanim}, N., {et~al.} 2016, \aap,
  586, A138

\bibitem[{{Redaelli} {et~al.}(2018){Redaelli}, {Bizzocchi}, {Caselli}, {Harju},
  {Chac{\'o}n-Tanarro}, {Leonardo}, \& {Dore}}]{Redaelli18}
{Redaelli}, E., {Bizzocchi}, L., {Caselli}, P., {et~al.} 2018, \aap, 617, A7

\bibitem[{{Redaelli} {et~al.}(2019){Redaelli}, {Bizzocchi}, {Caselli},
  {Sipil{\"a}}, {Lattanzi}, {Giuliano}, \& {Spezzano}}]{Redaelli19}
{Redaelli}, E., {Bizzocchi}, L., {Caselli}, P., {et~al.} 2019, \aap, 629, A15

\bibitem[{{Rudd} {et~al.}(1992){Rudd}, {Kim}, {Madison}, \& {Gay}}]{Rudd92}
{Rudd}, M.~E., {Kim}, Y.~K., {Madison}, D.~H., \& {Gay}, T.~J. 1992, Reviews of
  Modern Physics, 64, 441

\bibitem[{{Ruffle} {et~al.}(1999){Ruffle}, {Bettens}, {Terzieva}, \&
  {Herbst}}]{Ruffle99}
{Ruffle}, D.~P., {Bettens}, R.~P.~A., {Terzieva}, R., \& {Herbst}, E. 1999,
  \apj, 523, 678

\bibitem[{{Sabatini} {et~al.}(2020){Sabatini}, {Bovino}, {Giannetti},
  {Wyrowski}, {{\'O}rdenes}, {Pascale}, {Pillai}, {Wienen}, {Csengeri}, \&
  {Menten}}]{Sabatini20}
{Sabatini}, G., {Bovino}, S., {Giannetti}, A., {et~al.} 2020, \aap, 644, A34

\bibitem[{{Scherer} {et~al.}(2008){Scherer}, {Fichtner}, {Ferreira},
  {B{\"u}sching}, \& {Potgieter}}]{Scherer08}
{Scherer}, K., {Fichtner}, H., {Ferreira}, S.~E.~S., {B{\"u}sching}, I., \&
  {Potgieter}, M.~S. 2008, \apjl, 680, L105

\bibitem[{{Schlafly} {et~al.}(2014){Schlafly}, {Green}, {Finkbeiner}, {Rix},
  {Bell}, {Burgett}, {Chambers}, {Draper}, {Hodapp}, {Kaiser}, {Magnier},
  {Martin}, {Metcalfe}, {Price}, \& {Tonry}}]{Schlafly14}
{Schlafly}, E.~F., {Green}, G., {Finkbeiner}, D.~P., {et~al.} 2014, \apj, 786,
  29

\bibitem[{{Shirley}(2015)}]{Shirley15}
{Shirley}, Y.~L. 2015, Publications of the Astronomical Society of the Pacific,
  127, 299

\bibitem[{{Silsbee} {et~al.}(2018){Silsbee}, {Ivlev}, {Padovani}, \&
  {Caselli}}]{Silsbee18}
{Silsbee}, K., {Ivlev}, A.~V., {Padovani}, M., \& {Caselli}, P. 2018, \apj,
  863, 188

\bibitem[{{Silsbee} {et~al.}(2020){Silsbee}, {Ivlev}, {Sipil{\"a}}, {Caselli},
  \& {Zhao}}]{Silsbee20}
{Silsbee}, K., {Ivlev}, A.~V., {Sipil{\"a}}, O., {Caselli}, P., \& {Zhao}, B.
  2020, \aap, 641, A39

\bibitem[{{Sipil{\"a}} {et~al.}(2015{\natexlab{a}}){Sipil{\"a}}, {Caselli}, \&
  {Harju}}]{Sipila15}
{Sipil{\"a}}, O., {Caselli}, P., \& {Harju}, J. 2015{\natexlab{a}}, \aap, 578,
  A55

\bibitem[{{Sipil{\"a}} {et~al.}(2019){Sipil{\"a}}, {Caselli}, \&
  {Harju}}]{Sipila19b}
{Sipil{\"a}}, O., {Caselli}, P., \& {Harju}, J. 2019, \aap, 631, A63

\bibitem[{{Sipil{\"a}} {et~al.}(2015{\natexlab{b}}){Sipil{\"a}}, {Harju},
  {Caselli}, \& {Schlemmer}}]{Sipila15b}
{Sipil{\"a}}, O., {Harju}, J., {Caselli}, P., \& {Schlemmer}, S.
  2015{\natexlab{b}}, \aap, 581, A122

\bibitem[{{Spezzano} {et~al.}(2016){Spezzano}, {Bizzocchi}, {Caselli}, {Harju},
  \& {Br{\"u}nken}}]{Spezzano16}
{Spezzano}, S., {Bizzocchi}, L., {Caselli}, P., {Harju}, J., \& {Br{\"u}nken},
  S. 2016, \aap, 592, L11

\bibitem[{{Spitzer} \& {Tomasko}(1968)}]{Spitzer68}
{Spitzer}, Lyman, J. \& {Tomasko}, M.~G. 1968, \apj, 152, 971

\bibitem[{{Stone} {et~al.}(2019){Stone}, {Cummings}, {Heikkila}, \&
  {Lal}}]{Stone19}
{Stone}, E.~C., {Cummings}, A.~C., {Heikkila}, B.~C., \& {Lal}, N. 2019, Nature
  Astronomy, 3, 1013

\bibitem[{{van der Tak} \& {van Dishoeck}(2000)}]{VanDerTak00}
{van der Tak}, F.~F.~S. \& {van Dishoeck}, E.~F. 2000, \aap, 358, L79

\bibitem[{{Ward-Thompson} {et~al.}(1999){Ward-Thompson}, {Motte}, \&
  {Andre}}]{WardThompson99}
{Ward-Thompson}, D., {Motte}, F., \& {Andre}, P. 1999, \mnras, 305, 143

\bibitem[{{Wilson}(1999)}]{Wilson99}
{Wilson}, T.~L. 1999, Reports on Progress in Physics, 62, 143

\end{thebibliography}
%

\appendix
\section{CR ionisation rate and temperature of the models \label{DifferentT}}
\begin{figure}[!h]
\centering
\includegraphics[width=.5\textwidth]{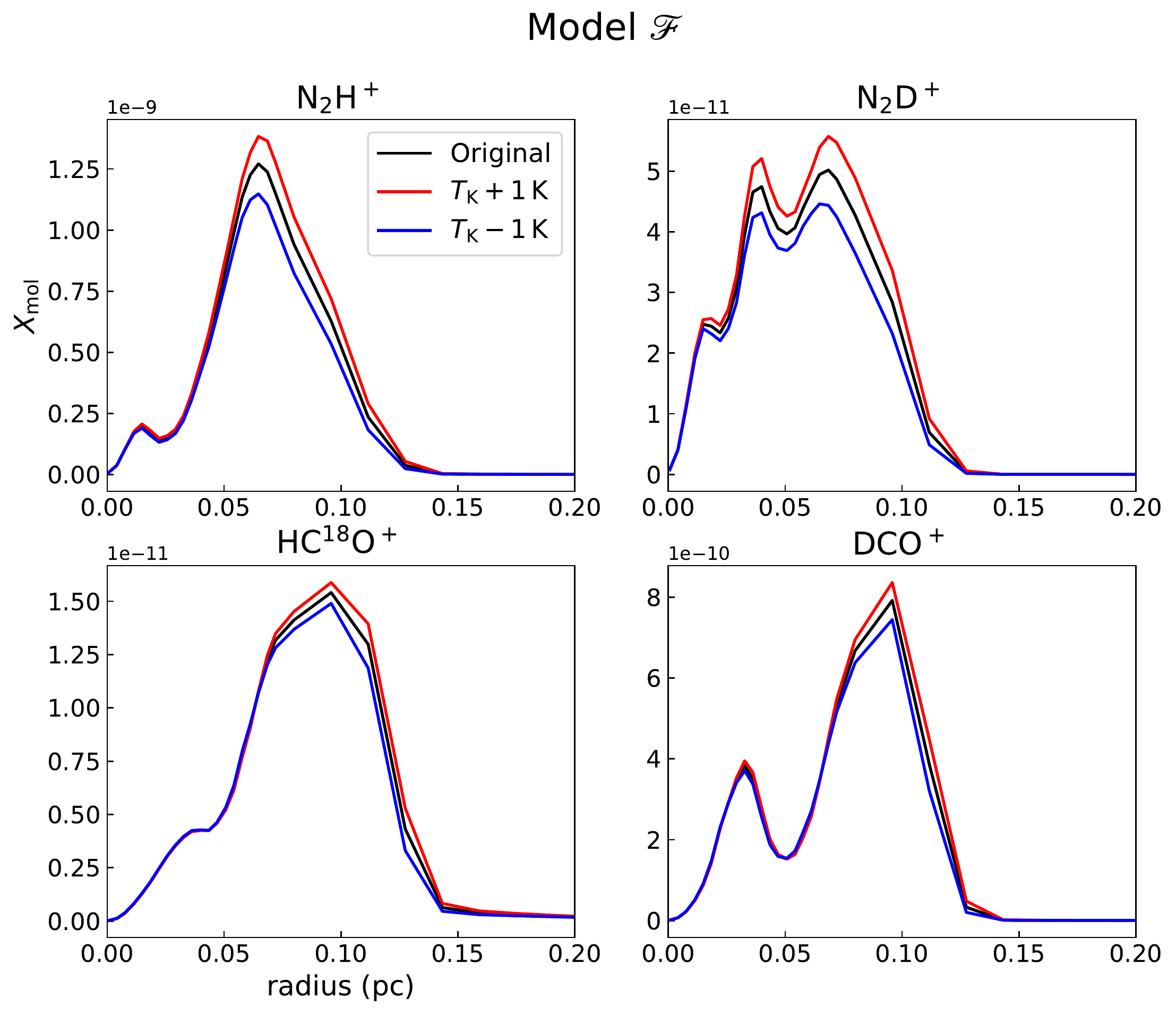}
\caption{Abundance profiles of \nnhp, \nndp, \hcop, and \dcop (from top-left to bottom-right panel) in the model \modf, using the original kinetic temperature profile from \cite{Keto15} (in black), and increasing or decreasing the temperature by $1\rm \, K$ (red and blue curves respectively). With respect to all other figures in this work, here we use the linear scale, since in the logarithmic scale the small variations would not be visible. The x-axis is limited to the central $0.2\, \rm pc$. \label{ModF_T}}
\end{figure}
\begin{figure}[!h]
\centering
\includegraphics[width=.5\textwidth]{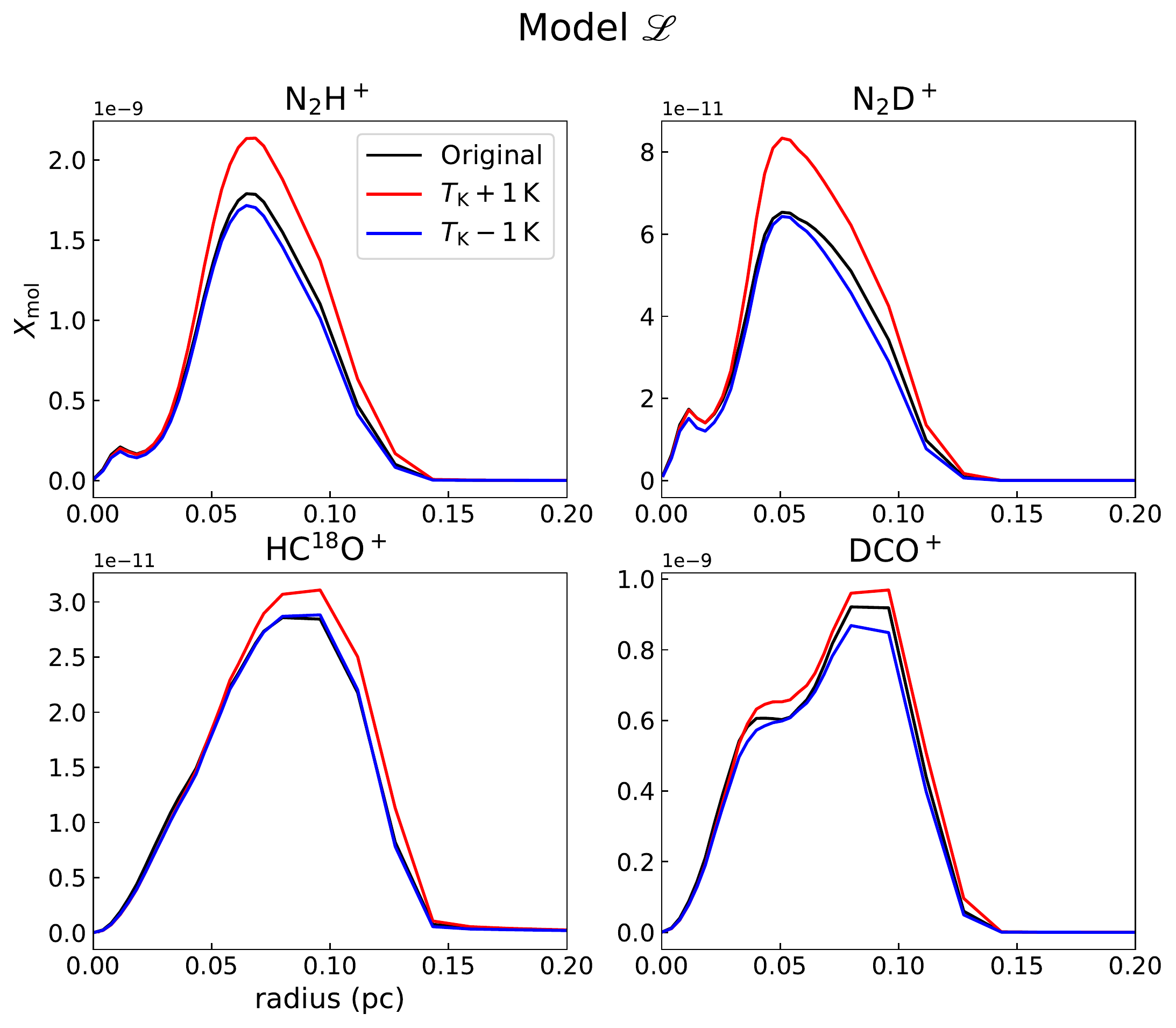}
\caption{Same as Fig. \ref{ModF_T}, but for model \modl. \label{ModL_T}}
\end{figure}

As mentioned in the main text, increasing (or decreasing) $\text{\crir}$ affects the gas temperature ($T_\mathrm{K}$). This, on the other hand, changes the excitation conditions for the different species. In principle one should self-consistently compute the temperature profiles (and the density profile) every time the CR ionisation rate is changed, which is beyond the scope of this work. Furthermore, we do not expect a significant temperature increase in the dense gas traced by the targeted molecular transitions, as the gas in this part of the core is mainly cooled by dust. This cooling mechanism is efficient enough that we do not expect important changes \cite[see also][]{Ivlev19}. A higher or lower \crir will affect mainly the temperature in the outer part of the core, where the kinetic temperature can be constrained for instance using ammonia inversion transitions (Schmiedeke et al. in prep.). 
\par
Nevertheless, in order to check how small temperature variations impact our radiative transfer modelling and conclusions, for the three models discussed in Sects. \ref{mod:1e6} and \ref{diff_times} (\modf, \modl, and \modh), we have run the chemical model in two more cases, obtained by increasing/decreasing the whole gas temperature profile uniformly by $1 \, \rm K$. \par
\begin{figure}[!h]
\centering
\includegraphics[width=.5\textwidth]{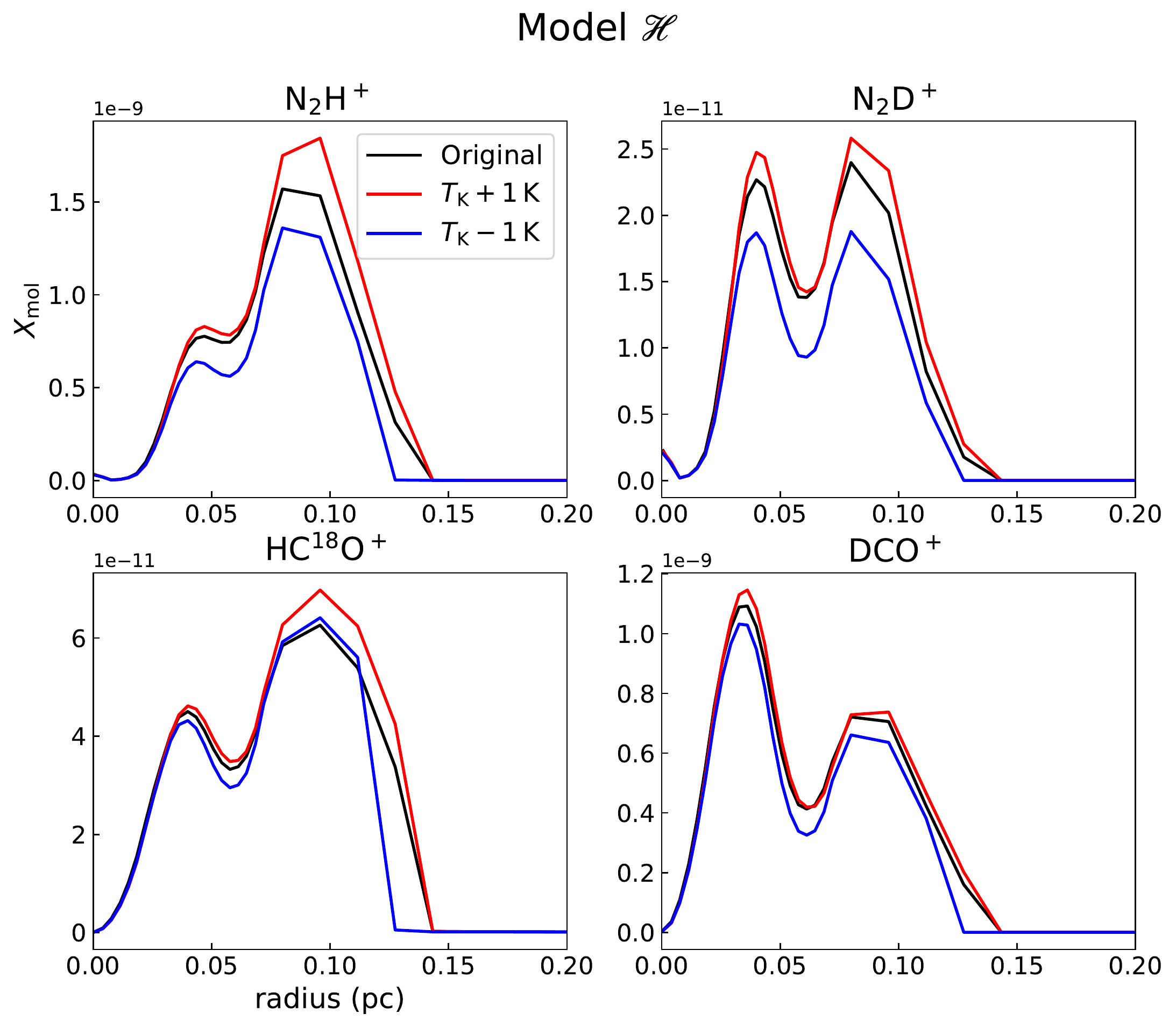}
\caption{Same as Fig. \ref{ModF_T}, but for model \modh. \label{ModH_T}}
\end{figure}

Figs. \ref{ModF_T} to \ref{ModH_T} show the resulting abundance profiles of the four species of interest obtained after changing the gas temperature. In general, an increase of the temperature leads to an increase of the abundance for all the molecules. The changes are however small, especially for \hcop ($4-6$\% on average) and \dcop ($6-7$\%), whilst they are higher for \nnhp ($11-13$\%) and \nndp ($12-18$\%). These small changes, together with the fact that varying the temperature in different models leads to similar abundance variations for a given molecule, leads us to conclude that taking into account these effects will not alter significantly our conclusions.

\section{Complete set of abundance profiles for models \modl, \modf, and \modh \label{AllModels}}
\begin{figure*}[!h]
\centering
\includegraphics[width=.9\textwidth]{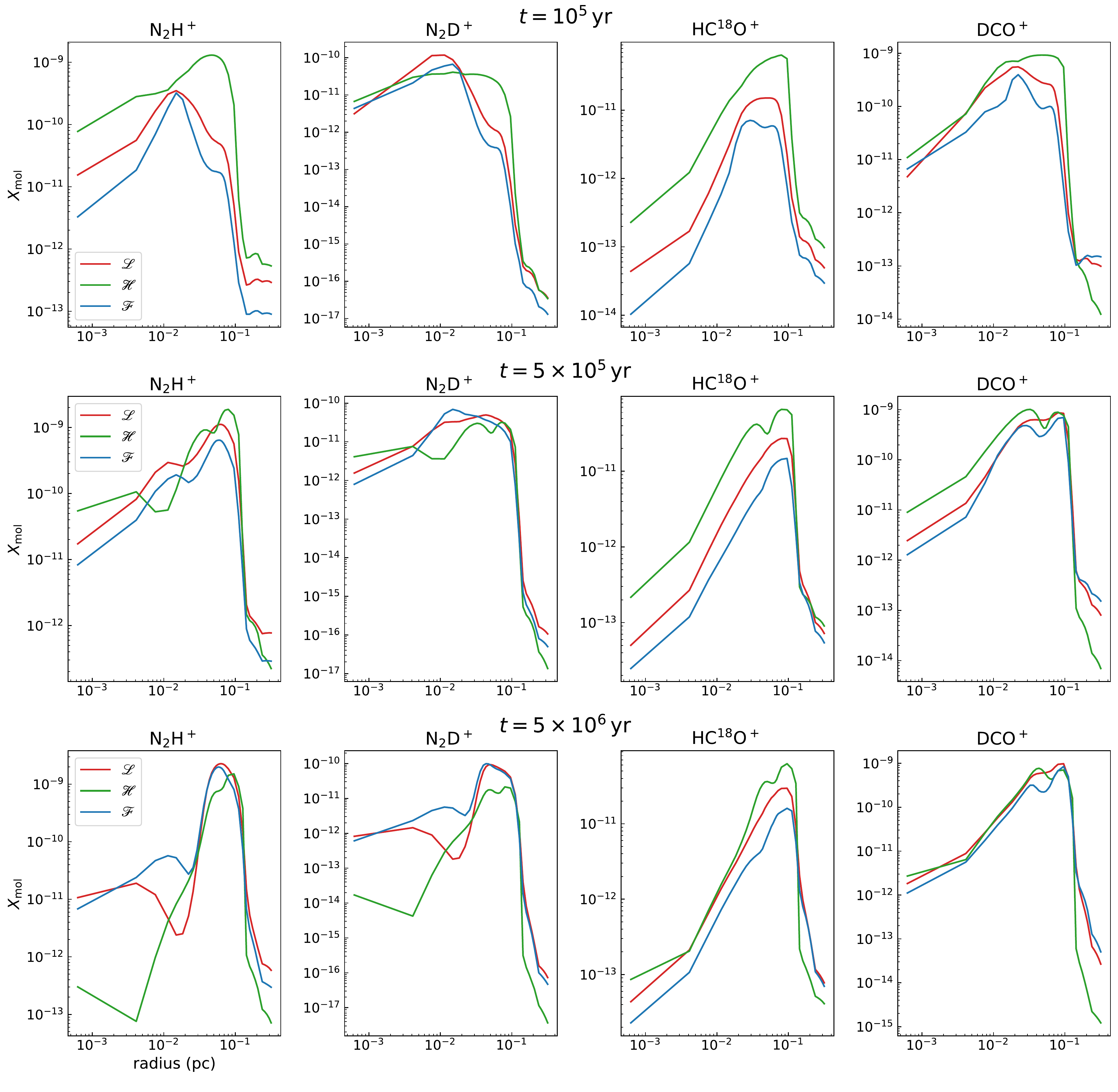}
\caption{Abundance profiles for the four analysed species (from left to right: \nnhp, \nndp, \hcdop, \dcop) at $t = 10^5$ (top row), $ t =  5 \times10^5\, \rm yr$ (middle row), and $ t =  5 \times10^6\, \rm yr$ (bottom row). The colours refer to different models used for the CR ionisation rate: the fiducial model \modf with $\text{\crir} = 1.3 \times 10^{-17} \, \rm s^{-1}$ is shown in blue, whilst the red and green curves show the models \modl and \modh  of PI18, respectively. \label{ChemMod_all}}
\end{figure*}

In Fig. \ref{ChemMod_all} we report the abundance profiles for the four targeted species obtained with model \modl, \modf, and \modh at the time steps not presented in Fig. \ref{ChemMod}.

\section{Uncertainties in the estimation of the \xab values \label{App:xab}}
\begin{figure}[!h]
\centering
\includegraphics[width=.9\textwidth]{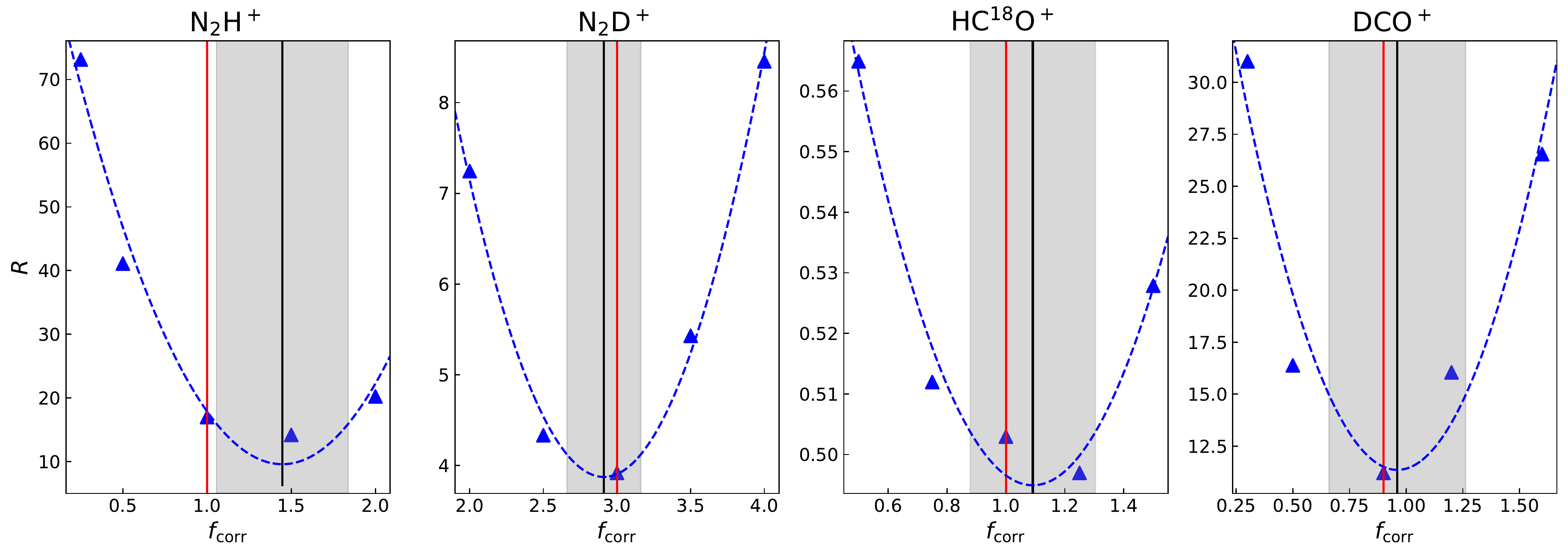}
\caption{Sum of the residuals, over all available transitions, computed with models obtained by changing the \xab value in the radiative transfer analysis. The synthetic spectra are produced using model \modl at $t=10^6\,\rm yr$ for all species but \nndp, for which we use the best-fit time step $t=5\times 10^5\,\rm yr$. The dashed, blue line shows the quadratic function fitted to the $R-f_\mathrm{corr}$ relation. In each panel, the vertical black line shows the $f \rm _{corr} ^{min}$ value that minimises the residuals, and the shaded grey area shows the corresponding $1\sigma$ interval. The \xab value evaluated in Sect. \ref{Analysis} is shown by the vertical red line. The panels refer to \nnhp, \nndp, \hcdop, and \dcop (from left to right). \label{ChiSquared}}
\end{figure}

As mentioned in Sect. \ref{mod:1e6}, the non-LTE radiative transfer code requires long computational times to converge, which prevents us to perform a full exploration of the parameter space, in particular to identify the best-fit \xab value for each species. Such a study is beyond the scope of this work, in which we focus on finding a general agreement between the distinct chemical models and the observed transitions. Finally, we highlight that this work is intrinsically characterised by large uncertainties, for instance in the physical model (L1544 is in reality not spherical), and in the chemical network (regarding e.g. rate coefficients and initial conditions).  Nevertheless, it is worth to verify with a more quantitative method if the approach used in the main text to find the \xab values is valid. 
\par
To this aim, we implemented a similar approach to that of \cite{Redaelli18}. Focusing on the results obtained for model \modl, we test five \xab factors around the value that provides the best-fit according to the analysis done in Sect. \ref{Analysis}. For each molecular species we then compute the residuals $R$ of the model with respect to the observations, following:
\begin{equation}
R =  \sum_{i,j} \left( T_\mathrm{MB,mod}^{i,j}(f_\mathrm{corr})  - T_\mathrm{MB, obs}^{i,j}\right)^2  \; ,
\end{equation}
where  $T_\mathrm{MB,mod/obs}^{i,j}$ represents the modelled/observed main beam temperature in the spectral channel $i$ for the rotational transition $j$. For each species we then fit a quadratic function to the $R-f_\mathrm{corr}$ relation. The $f\rm _{corr} ^{min}$ value that corresponds to the minimum value of the residuals is considered the best-fit value. Using standard error propagation from the uncertainties on the parameters of the quadratic relation, we estimate uncertainties on $f\rm _{corr} ^{min}$. \par
Figure \ref{ChiSquared} shows the results of this analysis. The $f\rm _{corr} ^{min}$ values are shown as black vertical lines, with the shaded area indicating the corresponding $\pm 1\sigma$ intervals. The \xab values found instead with the approach of Sect. \ref{Analysis} are shown as red vertical lines. For \nndp, \hcdop, and \dcop the differences between the two approaches are below 10\% and well below the uncertainty level. Only for \nnhp the two methods are not consistent. However, the discrepancy is only at a $1.2\sigma$ level. Furthermore, the value found ($f\rm _{corr} ^{min} = 1.45$) is still equal or smaller to the corrective factors found in all the other tested models. We conclude that, given the uncertainties, our method is robust and equivalent to a more quantitative, but time consuming, analysis.

\section{Cloud-evolution model: radiative transfer results \label{App:cloud_evo}}
\begin{figure*}[!h]
\centering
\includegraphics[width=.7\textwidth]{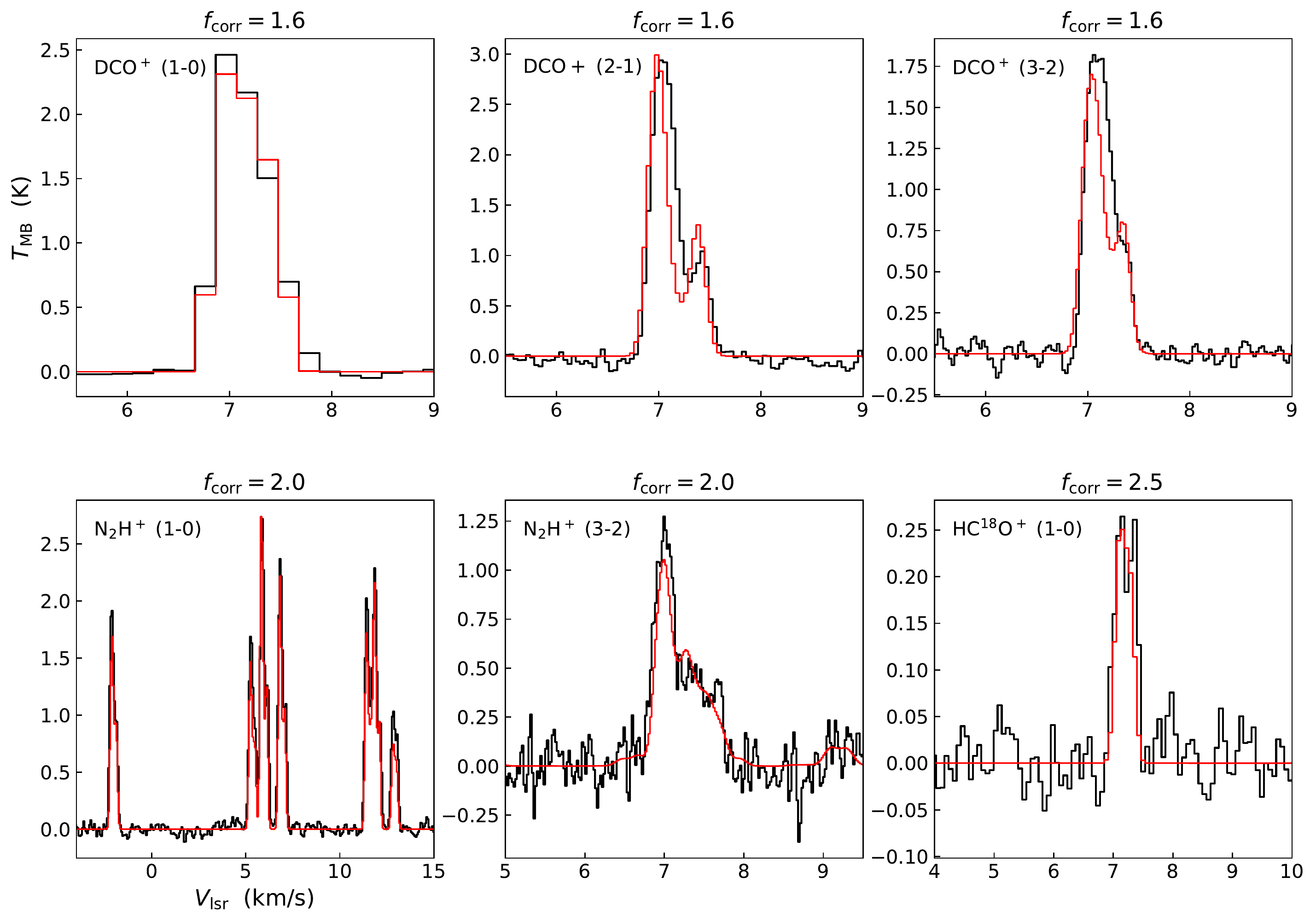}
\caption{Radiative transfer results (in red) overlaid to the observations (in black) for all the targeted transition obtained with the chemical model with initial cloud evolution to set the initial conditions, obtained at $t=10^6 \, \rm yr$ (see Sect. \ref{Sec:CloudEvo}). The model implements constant \crir equal to the fiducial model ($\text{\crir} = 1.3 \times 10^{-17} \, \rm s^{-1}$). The top row shows the \dcop lines, while the bottom row presents the two \nnhp transitions and the \hcdop (1-0) line, from left to right. The corresponding panels for \nndp are presented in Fig. \ref{n2dp_modCE}. The used  \xab value is shown on top of each panel. \label{CE_mod}}
\end{figure*}

In this appendix we report the radiative transfer results obtained from the model with initial cloud evolution and constant $\zeta_2 = 1.3 \times 10^{-17} \rm \, s^{-1}$. The modelled spectra are shown in Fig. \ref{CE_mod}. It can be seen that, in general, this model requires equal or higher corrective factors with respect for instance to model \modl or model \modf.

\section{Model with $\text{\crir} = 3 \times 10^{-17} \rm \, s^{-1}$: resulting line profiles. 
\label{App:enhanchedMod}}

\begin{figure*}[!h]
\centering
\includegraphics[width=.7\textwidth]{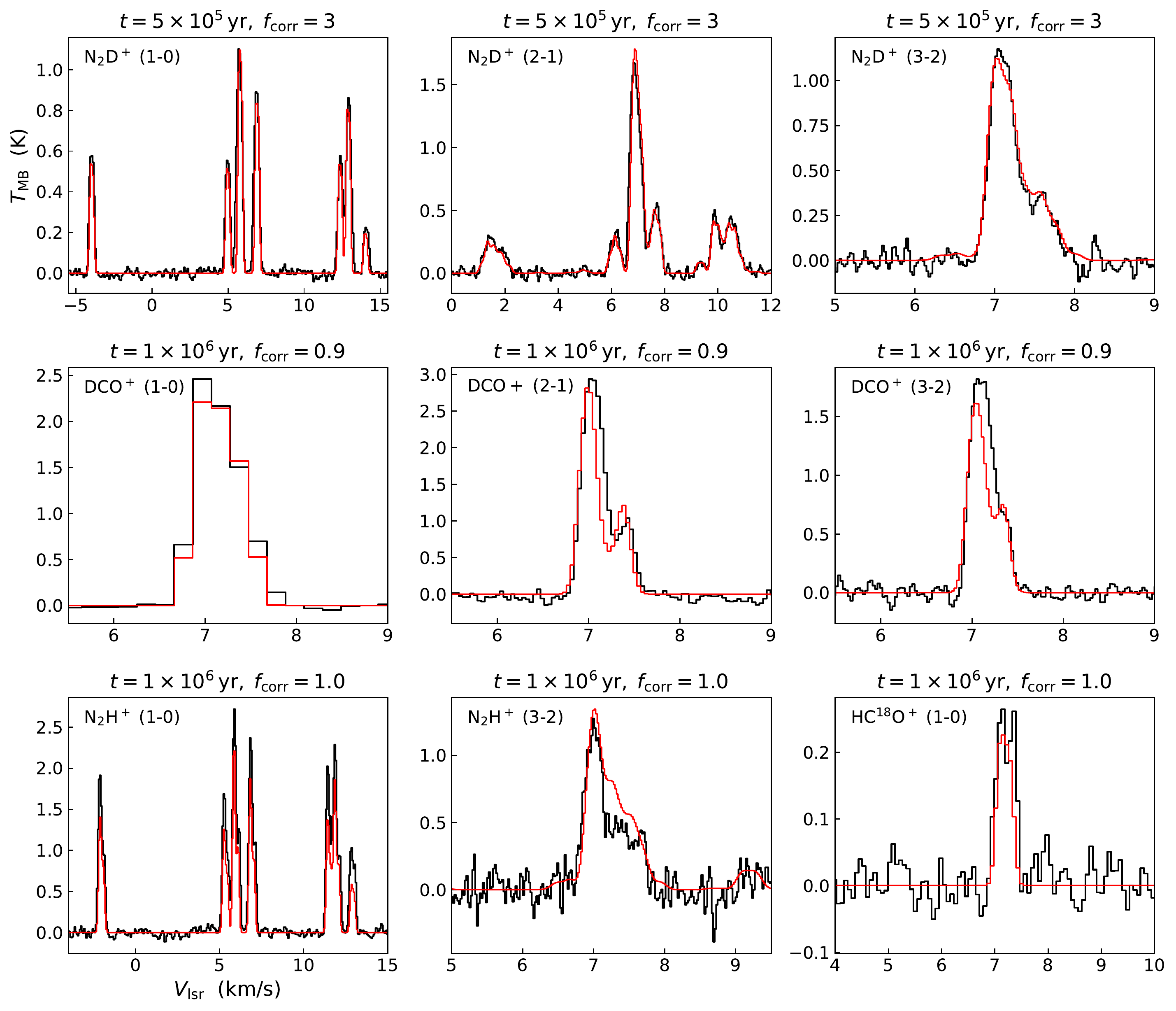}
\caption{Radiative transfer results (in red) overlaid to the observations (in black) for all the targeted transition obtained with the chemical model with constant $\text{\crir} = 3 \times 10^{-17} \rm \, s^{-1}$, using the same evolutionary time steps and \xab factors that provide the best fit for the fiducial model \modf. The top row shown the \nndp line, the middle row shows the \dcop lines, and the bottom row presents the two \nnhp transitions and the \hcdop (1-0) line, from left to right. The used time step and \xab value are shown in the title of each panel. \label{Enhanced_mod}}
\end{figure*}

Figure \ref{Enhanced_mod} shows the results of the radiative transfer applied to the chemical model with constant $\text{\crir} = 3 \times 10^{-17} \rm \, s^{-1}$.
\end{document}